\documentclass[a4paper,twocolumn,unpublished]{quantumarticle}

\pdfoutput=1

\usepackage{caption}
\usepackage{subcaption}

\usepackage{float}
\usepackage[utf8]{inputenc}
\usepackage[english]{babel}
\usepackage[T1]{fontenc}
\usepackage{amsmath}
\usepackage{amssymb}
\usepackage{hyperref}
\usepackage{tikz}
\usepackage{lipsum}

\usepackage[linesnumbered,ruled]{algorithm2e}

\newcommand{\IQP}{\textbf{IQP }}
\newcommand{\HSC}{\textbf{HSC }}
\newcommand{\RND}{\textbf{RND }}

\newcommand{\estimate}[1]{\tilde{#1}}

\begin{document}

\title{Bidirectional Path Integral Monte Carlo Simulation of Quantum Circuits}

\author{Luís Paulo Santos}
\email{psantos@di.uminho.pt}
\affiliation{HASLab, INESC TEC / Universidade do Minho / INL, Portugal}
\homepage{https://luisps.github.io/}

\author{Thomas Bashford-Rogers}
\email{Thomas.Bashford-Rogers@warwick.ac.uk}
\affiliation{Warwick Manufacturing Group, Warwick University, Coventry, United Kingdom}
\homepage{https://thomasbashfordrogers.com/}

\begin{abstract}
We propose an approximate classical algorithm for estimating quantum circuits' transition amplitudes, $\langle \Psi_f | \hat{U} | \Psi_0 \rangle$, based on Monte Carlo sums over bidirectionally sampled Feynman paths. The algorithm generates Feynman paths, i.e., sequences of intermediate computational basis states, by combining information from random walks initiated at both the input and output states, $\Psi_0$ and $\Psi_f$, respectively. It further combines this bidirectional approach with Multiple Importance Sampling, a powerful variance reduction technique that avoids variance peaks caused by divisions by the very low sampling probabilities associated with hard to reach paths.
Empirical data demonstrate the superior convergence rate exhibited by the bidirectional algorithm over a unidirectional, forward-only sampling approach. This superiority is due to the effectiveness of the proposed algorithm in finding and weighting paths with non-zero amplitude in an extremely sparse path space. 
Sum of paths algorithms exhibit very favourable space complexity. Monte Carlo replaces exhaustive enumeration of the exponentially large path space by stochastic sampling, the computational cost being determined by the variance of the estimator. This work demonstrates these two features by presenting results on very large quantum circuits, with up to 4096 qubits.
\end{abstract}

\maketitle

\section{Introduction}
\label{sec:Introduction}

Current quantum computing platforms are scarce and noisy, with performance limited by decoherence. Execution of quantum circuits is therefore expensive, due to scarcity, and limited to shallow circuits due to short decoherence times. Classical simulation of quantum circuits is a fundamental capability enabling researchers and practitioners to validate algorithms, study quantum advantage, and benchmark quantum solutions. Simulating arbitrary circuits requires time and/or space that grow exponentially with circuit size, given by the number of qubits $n$ and gate depth $L$, motivating research efforts to efficiently simulate families of quantum circuits. These techniques exploit characteristics exhibited by most quantum circuits, such as structure, redundancy, and low magic, among others.

This paper proposes an approximate classical algorithm for estimating quantum circuits transition amplitudes, $\langle \Psi_f | \hat{U} | \Psi_0 \rangle$, based on Monte Carlo sums over bidirectionally sampled Feynman paths. The algorithm is inspired by current practice in physically based global illumination. Bidirectional Path Tracing (BDPT) is a Monte Carlo algorithm \cite{Lafortune1998, Veach1997} that excels at finding low-probability light paths connecting the light source to the camera. Similarly, this work shows that a bidirectional approach to the Path Integral Monte Carlo (PIMC) problem, within the context of simulating a quantum circuit described by a time-evolution operator $\hat{U}$,  efficiently finds low-probability paths connecting $|\Psi_0\rangle$ to $|\Psi_f\rangle$. Challenges raised by the sparsity of the Feynman path space are efficiently handled by the bidirectional sampling procedure.

Simulation algorithms based on summing over Feynman paths allow for the classical evaluation of transition amplitudes $\langle \Psi_f | \hat{U} | \Psi_0 \rangle$ with $\mathcal{O}(nL)$ space complexity, since the simulator only maintains data on the circuit itself and a running sum over all Feynman paths' amplitudes. The corresponding path space, $\mathcal{P}$, has cardinality $|\mathcal{P}| = 2^{n*(L-1)}$. The exponential advantage in space complexity, compared to Schrödinger-based approaches, comes at the cost of increased time complexity associated with $\mathcal{P}$'s exponentially large number of paths.

Monte Carlo (MC) techniques evaluate the estimator  $\tilde{I} \approx \langle \Psi_f | \hat{U} | \Psi_0 \rangle$ using $N$ stochastically generated samples, each sample being a classical path from $|\Psi_0 \rangle$ to $|\Psi_f \rangle$, $\bar{z}_i \in \mathcal{P}, \ i=0 \ldots N-1$.  The goal is to overcome the curse of dimensionality problem by evaluating many fewer samples than the total number of paths in the path space, $N \ll |\mathcal{P}|$. 

For a finite number of samples, the MC estimator will exhibit variance, whose magnitude depends on three major factors: i) how effectively contributing paths, $f(\bar{z}_i)\neq 0$, are found within a often very sparse path space, ii) how well samples' contributions, $f(\bar{z}_i)$, are weighted, and iii) destructive interference among paths with mutually cancelling contributions, referred to as the numerical sign problem. 

The proposed Monte Carlo estimator improves the variance associated with path space sparseness and paths' weighting, thus addressing items i) and ii) above. The bidirectional approach samples path space by combining random walks initiated from both $|\Psi_0 \rangle$ and $|\Psi_f \rangle$, thus increasing the probability of finding non-zero paths within a highly sparse domain. The paths' contributions are weighted using Multiple Importance Sampling (MIS), further improving the MC convergence rate by avoiding divisions by the very small probabilities associated with hard to find paths in $\mathcal{P}$.

This paper details the proposed bidirectional sampling (BS) approach and provides empirical evidence supporting the following three hypotheses:
\begin{description}
    \item[H1 -- Convergence rate: ] BS achieves better convergence than a unidirectional forward sampling algorithm.  
    \item[H2 -- Simulation of large circuits: ] BS allows estimating transition amplitudes for very large quantum circuits (e.g., 4096 qubits) because its memory complexity is $\mathcal{O}(nL)$ and it efficiently finds non-zero paths in a very large and sparse path space. 
    \item[H3 -- Sparsity and numerical sign problem: ] BS improvements in the convergence rate, over the unidirectional approach, are due to a more efficient sampling of the sparse path space, rather than to a better handling of destructive interference, i.e., MC's numerical sign problem.
\end{description}

The document is structured as follows. The next section presents the most important related work; section \ref{sec:PIMC} reviews Monte Carlo integration, bidirectional and multiple importance sampling. Section \ref{sec:BDMC} describes the proposed approach, the following section describes the experimental methodology and section \ref{sec:Results} analyses experimental results. Section \ref{sec:conclusions} concludes the paper.

\section{Related work}

Classical simulation of quantum circuits is a central problem in quantum computing, underpinning algorithm validation, benchmarking of near-term devices, and the study of quantum advantage. The computational cost of simulating a quantum system typically  grows exponentially with the number of qubits in the worst case \cite{Bernstein1997}, motivating a diverse set of simulation techniques that exploit structure, approximations, redundancy, low magic, or alternative representations. Simulation methods can be broadly categorized by the representation of quantum mechanics they are based upon, including the Schrödinger, Feynman, and Heisenberg perspectives.

\subsection{The Schrödinger Picture}

In the Schrödinger picture, the quantum state evolves under the action of time evolution unitary operators $\hat{U}$. Simulation methods in this paradigm explicitly represent or approximate the {\bf full state vector} $| \Psi \rangle$ of size $2^n$, updating it gate by gate \cite{Nielsen2010}. Single- and two-qubit gates correspond to sparse linear transformations $\hat{U}_g$ applied to $| \Psi \rangle$. This is a conceptually simple and exact approach. Memory requirements, however, grow exponentially with system size ($\mathcal{O}(2^n)$), limiting simulations to roughly 30–50 qubits depending on available resources. 

{\bf Tensor network} approaches exploit the structure inherent in limited-entanglement quantum circuits \cite{Biamonte2017,Markov2008}. The circuit is mapped to a network of tensors, and the simulation reduces to contracting this network. The cost depends on the complexity of the contraction rather than the number of qubits. Tensor networks enable the simulation of systems that are significantly larger than those possible with state-vector approaches. Common techniques include matrix product states (MPS), projected entangled pair states (PEPS), and more general contraction schemes optimized via graph partitioning \cite{Cirac2021}. Highly entangled circuits, such as random circuits, require large intermediate tensors, bringing back exponential requirements.

{\bf Decision diagrams} represent quantum states and operators as compressed graph structures that exploit redundancy and regularity in amplitudes \cite{Wille2023, Hong2025}. Operations correspond to graph manipulations, often yielding substantial memory savings. Efficiency depends critically on the amount of structure present; highly random or chaotic circuits degrade compression, leading to exponential growth. Nevertheless, decision diagrams provide a powerful alternative when the quantum circuits exhibit repeating patterns or symmetries.

\subsection{Feynman Path Integral Methods}

In the Feynman picture, amplitudes are expressed as {\bf sums over computational paths}. Each path $\bar{z}=(s_0=\Psi_0, s_1, \ldots , s_L=\Psi_f)$, with length $L$, corresponds to a sequence of intermediate basis states and the total amplitude is obtained by summing contributions from all paths $\bar{z} \in \mathcal{P}$:
\begin{widetext}
\begin{equation}    
\langle \Psi_f | \hat{U} | \Psi_0 \rangle = \sum_{\bar{z} \in \mathcal{P}} f(\bar{z}) = \sum_{(s_1, \ldots , s_{L-1}) \in \mathcal{P}} \left[ \langle \Psi_{f} | \hat{U}_{L-1} | s_{L-1} \rangle \prod_{t=L-2}^{1} \left( \langle s_{t+1} | \hat{U}_t | s_t \rangle \right) \  \langle s_1 | \hat{U}_0 | \Psi_0 \rangle \right] .
\label{eq:FeynmanSum}
\end{equation}
\end{widetext}

This formulation avoids storing the full state vector but incurs a combinatorial explosion in the number of paths \cite{Aaronson2017, Burgholzer2021}. Practical algorithms rely on path pruning or on hybrid Schrödinger–Feynman methods, in which the circuit is partitioned into subcircuits that are simulated independently and recombined. Exact Feynman-based approaches are particularly effective for circuits with small width or for simulations that trade memory for time. They underpin some of the most powerful classical simulations of random quantum circuits used in benchmarking quantum advantage experiments.

\subsection{Heisenberg Picture Methods}

In the Heisenberg picture, operators evolve backward through the circuit while the state remains fixed. A prominent approach is {\bf Pauli propagation} \cite{Rall2019}. Observables are decomposed into sums of Pauli operators, and each Pauli term is propagated through the circuit using conjugation rules. For Clifford circuits, Pauli operators remain Pauli under conjugation, enabling efficient simulation via stabilizer techniques. For general circuits, non-Clifford gates cause Pauli operators to expand into linear combinations, leading to exponential growth. Truncation strategies that yield approximate estimates were proposed to mitigate exponential growth \cite{Rudolph2026}. Pauli-based methods are particularly useful for estimating expectation values.

{\bf Stabilizer simulation} tableau-based methods constitute the canonical Heisenberg approach for Clifford circuits \cite{aaronson2004,gottesman1998}. Operators are represented implicitly via stabilizer generators, with evolution corresponding to symplectic updates of the tableau. The algorithm tracks how a basis of Pauli operators transforms under conjugation, rather than tracking a single observable.

{\bf Pauli Transfer Matrix} approaches propagate operators using a linear map in operator space \cite{Greenbaum2015}. Representing quantum channels as matrices acting on the Pauli basis makes these algorithms particularly suited for simulating noisy circuits. The {\bf Matrix Product Operator} (MPO) Heisenberg Evolution is the Heisenberg analog of tensor networks, with observables represented as MPOs \cite{Prosen2007}. The rationale is to exploit the fact that operator entanglement often, but not always, grows slower than state entanglement.

\subsection{Monte Carlo and Approximate Simulation}

Monte Carlo methods provide a framework for approximate simulation, estimating amplitudes or expectation values via stochastic sampling. General PIMC can be broadly described as stochastically sampling $N$ paths $\bar{z}_i$, with probability $p(\bar{z}_i)$, over path space $\mathcal{P}$, evaluating samples' values $f(\bar{z}_i)$, and linearly combining these with weights $\frac{1}{p(\bar{z}_i)}$. The PIMC estimator, $\tilde I _{MC}$, takes the form 
\begin{equation}
    \langle \Psi_f | \hat{U} | \Psi_0 \rangle \approx \tilde I_{MC} = \frac{1}{N}\sum_{i=0}^{N-1} \frac{f(\bar{z}_i)}{p(\bar{z}_i)}.
    \label{eq:MCestimator}
\end{equation} 

Typical approaches {\bf importance sample Feynman paths} according to some probability distribution, yielding unbiased estimators with variances that depend on the paths' interference structure \cite{Sauer2001,Temme2017}. Alternative approaches use {\bf quasi-probability decompositions}, expressing non-Clifford gates as linear combinations of Clifford operations with signed weights \cite{Pashayan2015, Bravyi2019}. Sampling over these decompositions enables simulation using efficient stabilizer methods, at the cost of a variance overhead quantified by the negativity of the Clifford weights. Hybrid methods combine Monte Carlo with tensor networks or decision diagrams, sampling over contraction orders or diagram branches \cite{Ferris2012}. These approaches can significantly extend the reach of classical simulation, particularly for circuits with moderate non-Clifford content (i.e., low magic). A key challenge is controlling variance, which can grow exponentially with the circuit's depth, non-classicality, and, most importantly, with the {\bf numerical sign problem} that emerges from destructive interference among paths \cite{Ayral2025}. Techniques such as importance sampling, control variates, and stratified sampling have been proposed to improve efficiency \cite{Kalos1964}, but are ineffective against the exponential sampling requirements imposed by destructive interference patterns. 

\subsection{Simulation Hardness}
\label{subsec:hardness}

Classical simulation of quantum circuits is widely believed to be intractable in general, meaning that it requires resources (space and/or time) that grow exponentially with circuit size. The metrics used to quantify this hardness depend on the underlying algorithm: the efficiency of the different simulation paradigms depends on distinct structural properties of the quantum state or circuit. Accordingly, a variety of metrics has been proposed, each capturing the challenges faced by a specific class of simulators.

Vanilla full-state Schrödinger approaches' cost scales as $\mathcal{O}(2^n)$ in both space and time. This is a worst-case metric that does not capture potential improvements from exploiting the quantum circuit's structure. Low-entanglement states admit efficient representations (e.g., matrix product states), suggesting that bipartite entropies or the maximum Schmidt rank can serve as proxies for simulation efficiency in tensor network approaches \cite{Vidal2003, Markov2008, Orus2014}. Entanglement alone, however, does not characterize classical hardness: stabilizer states can exhibit high entanglement while remaining efficiently simulable due to their strong algebraic structure. Additional resources -- such as magic, phase complexity, or lack of redundancy -- are necessary to capture the full computational difficulty. Additionally, treewidth -- a measure of how well the tensor network graph can be decomposed into a tree of small overlapping clusters -- is often used to characterize contraction complexity.

Decision diagram simulators exploit redundancy by sharing identical substructures. Their complexity is measured by the compression ratio, $\frac{2^n}{N_{\mathrm{DD}}}$, with $N_{\mathrm{DD}}$ the number of nodes in the diagram. It quantifies the redundancy of the amplitude tensor, i.e., the number of distinct subtensors in its recursive decomposition \cite{Viamontes2009,burgholzer2020}. Highly structured states (e.g., stabilizer states) exhibit strong compression, whereas generic states lead to $N_{\mathrm{DD}} \sim 2^n$.

Clifford circuits can be efficiently simulated using stabilizer formalisms. The challenge is the presence of non-Clifford resources, quantified by magic, often defined as the stabilizer extent:
\begin{equation}
    \xi(\psi) = \min \left(\sum_k |c_k|\right)^2 \quad \text{, } |\psi\rangle = \sum_k c_k |\phi_k\rangle,
\end{equation}
with $|\phi_k\rangle$ being stabilizer states. If $|\psi\rangle$ is a stabilizer state, then $\xi = 1$. Otherwise, $\xi$ grows (often exponentially) with the number of T gates. Since the simulation cost in many stabilizer decomposition algorithms scales polynomially with $\xi$, it is a key hardness metric \cite{Leone2022}. Additional measures of magic include robustness of magic \cite{Howard2017} and mana \cite{Veitch2012}.  

Approaches based on the sum of Feynman paths express the amplitude $\langle \psi_f |\, \hat{U} \,|\psi_0 \rangle$ as a sum over exponentially many computational paths. Hardness arises from the number of contributing paths (i.e., path space density) and the lack of degeneracy, i.e., few paths share identical contributions \cite{Dawson2006}. This connects directly to redundancy. If many paths have identical, eventually zero, contributions, they can be grouped; otherwise, the sum is irreducible.

The complexity of Heisenberg-based methods depends on the growth of the Pauli support and on the number of Pauli terms generated under conjugation as observables are propagated backward through the circuit \cite{Garcia2012}. Clifford circuits preserve Pauli structure, but non-Clifford gates cause a proliferation of terms. This growth is closely related to magic and can be understood as its operator-space analog.

The efficiency of Path Integral Monte Carlo methods depends on two key factors \cite{Troyer2005,Temme2017,Pashayan2015}: how effectively the method finds contributing paths in the highly sparse path space, and how hard the numerical sign problem is, caused by destructive interference among paths. The latter is often quantified by resorting to the inverse squared average phase, $\chi_{\mathcal{P}} = \langle e^{i\phi} \rangle_\mathcal{P} ^{-2}$ \cite{Alexandru2018,Xu2022}.

The amplitude of a path $\bar{z}$ can be written as $f(\bar{z}) = |f(\bar{z})| e^{i\phi_{\bar{z}}}$; trivially, the respective phase is given by $e^{i\phi_{\bar{z}}} = \frac{f(\bar{z})}{|f(\bar{z})|}$. $\mathcal{P}$'s average phase is defined as 
\begin{equation}
    \langle e^{i \theta} \rangle_{\mathcal{P}} = \frac{|\sum_{\bar{z} \in \mathcal{P}}f(\bar{z})|}{\sum_{\bar{z} \in \mathcal{P}}|f(\bar{z})|} \in [0 \ldots 1].
    \label{eq:averagePhase}
\end{equation}
 If the phases of contributing paths oscillate, then the numerator above becomes small due to destructive interference and $\langle e^{i \theta} \rangle_{\mathcal{P}}$ approaches $0$, indicating a hard numerical sign problem.

The inverse squared average phase \cite{Alexandru2022} is given by
\begin{equation}    
\chi_{\mathcal{P}} = \langle e^{i\phi} \rangle_\mathcal{P} ^{-2} = \left( \frac{\sum_{\bar{z} \in \mathcal{P}}|f(\bar{z})|}{|\sum_{\bar{z} \in \mathcal{P}} f({\bar{z}})|} \right)^2 \in [1 \ldots \infty].
\label{eq:chi}
\end{equation}
$\chi_{\mathcal{P}}$ is often used to quantify the Monte Carlo simulation hardness because for highly oscillatory path spaces it is closely related to the estimator's relative variance:
\begin{equation}
RVar( \tilde{I}_{MC}) = \frac{Var(\tilde{I}_{MC})}{ (\mathbb{E}[\tilde{I}_{MC}])^2} \approx \frac{1}{N \langle e^{i \theta} \rangle^2_{\mathcal{P}}} = \frac{\chi_\mathcal{P}}{N}
\label{eq:RVar}
\end{equation}
For fixed relative variance and highly oscillatory path spaces, sampling requirements scale as $N \propto \mathcal{O}(\langle e^{i \theta} \rangle^{-2}_{\mathcal{P}}) = \mathcal{O}(\chi_\mathcal{P})$.

These metrics are deeply connected. Magic can be thought to induce phase randomness, destroying redundancy in the amplitude tensor, increasing Pauli complexity, and leading to an exponentially large inverse-squared average phase. 



\section{Monte Carlo Path Integration}
\label{sec:PIMC}

A path integral is an integral over a space of functions, which are referred to as paths, trajectories, histories, etc., depending on the context. The path integral is typically denoted by
\begin{equation}
    I = \int_{\mathcal{P}} F(\bar{z}(x)) \, d\mu(\bar{z}(x))
\end{equation}
where $\mathcal{P}$ is the space of paths $\bar{z}(x)$, $F(\bar{z}(x))$ is a functional that assigns a value to the entire path, and $d\mu(\bar{z})$ is a measure over $\mathcal{P}$. In the context of this work, we consider fixed-length discretized paths consisting of $L$ transitions and represented by a sequence of $L + 1$ states $(x_0, x_1, \ldots , x_{L-1}, x_{L})$. The corresponding integral over paths of length $L$ is

\begin{equation}
    I_L = \int_{\mathcal{P}_L} F(x_0, \ldots , x_L) \, d x_0 \ldots  d x_L .
    \label{eq:fixedlengthintegral}
\end{equation}


Quantum circuits already entail a discretization of the quantum state time evolution into $L$ time intervals, where $L$ is the circuit's depth. The problem being addressed is therefore of the form depicted by equation \ref{eq:fixedlengthintegral}.

Computing $\langle \Psi_f | \hat{U} | \Psi_0 \rangle$ entails integrating over the path space 
{\small
\begin{equation}
    \mathcal{P}_L = \{(s_0, \ldots , s_L) : s_0=\Psi_0, s_L=\Psi_f , s_{1 \ldots L-1} \in \mathcal{S} \}
\end{equation}
} 
with $\mathcal{S}$ being the set of all possible states at each vertex along the path. This work integrates over paths defined in the computational basis, therefore $\mathcal{S}=\{0 \ldots 2^n-1\}$ is discrete, with $n$ being the number of qubits in the circuit. Given that there are $L-1$ intermediate states (or path vertices, excluding $s_0$ and $s_L$), then $\mathcal{P} = \mathcal{S}^{L-1}$ and $|\mathcal{P}| = 2^{n*(L-1)}$. A path is therefore a classical object, denoting a history of the system evolution over classical states along $L$ time intervals, each corresponding to a layer of quantum gates.

The transition amplitude is given by a sum:
\begin{equation}
    \langle \Psi_f | \hat{U} | \Psi_0 \rangle = I = \sum_{\bar{z}_i \in \mathcal{P}} f(\bar{z}_i),
    \label{eq:sumPaths}
\end{equation}
and $f(\bar{z}_i): \mathcal{P} \rightarrow \mathbb{C}$ is evaluated according to the summand in equation \ref{eq:FeynmanSum}.

Monte Carlo integration enables efficient estimation of equation \ref{eq:sumPaths} by stochastically drawing $N \ll |\mathcal{P}|$ samples (paths) from $\mathcal{P}$, as given by:
\begin{equation}
    I = \sum_{\bar{z}_i \in \mathcal{P}} f(\bar{z}_i) \approx \tilde{I}_{MC} = \frac{1}{N} \sum_{i=0}^{N-1}  \frac{f(\bar{z}_i)}{p(\bar{z}_i)} \ ,
    \label{eq:MCestimation}
\end{equation}
where $p(\bar{z}_i)$ is the probability with which the path $\bar{z}_i$ is sampled. The variance of the estimator is given by
\begin{equation}
    Var(\tilde{I}_{MC}) = \frac{1}{N} \left( \sum_{i=0}^{N-1}  \frac{f(\bar{z}_i)^2}{p(\bar{z}_i)} - I^2 \right)
\end{equation}
Since the standard error is the square root of the variance, it decreases with the reciprocal of the square root of the number of samples: $\sigma_{MC} = \sqrt{Var(\tilde{I}_{MC})} \approx \mathcal{O}(\frac{1}{\sqrt{N}}) $.

\subsection{Forward Importance Sampling}
\label{subsec:PIMC-FIS}

\begin{figure}[htb]
    \centering
\includegraphics[width=0.9\columnwidth]{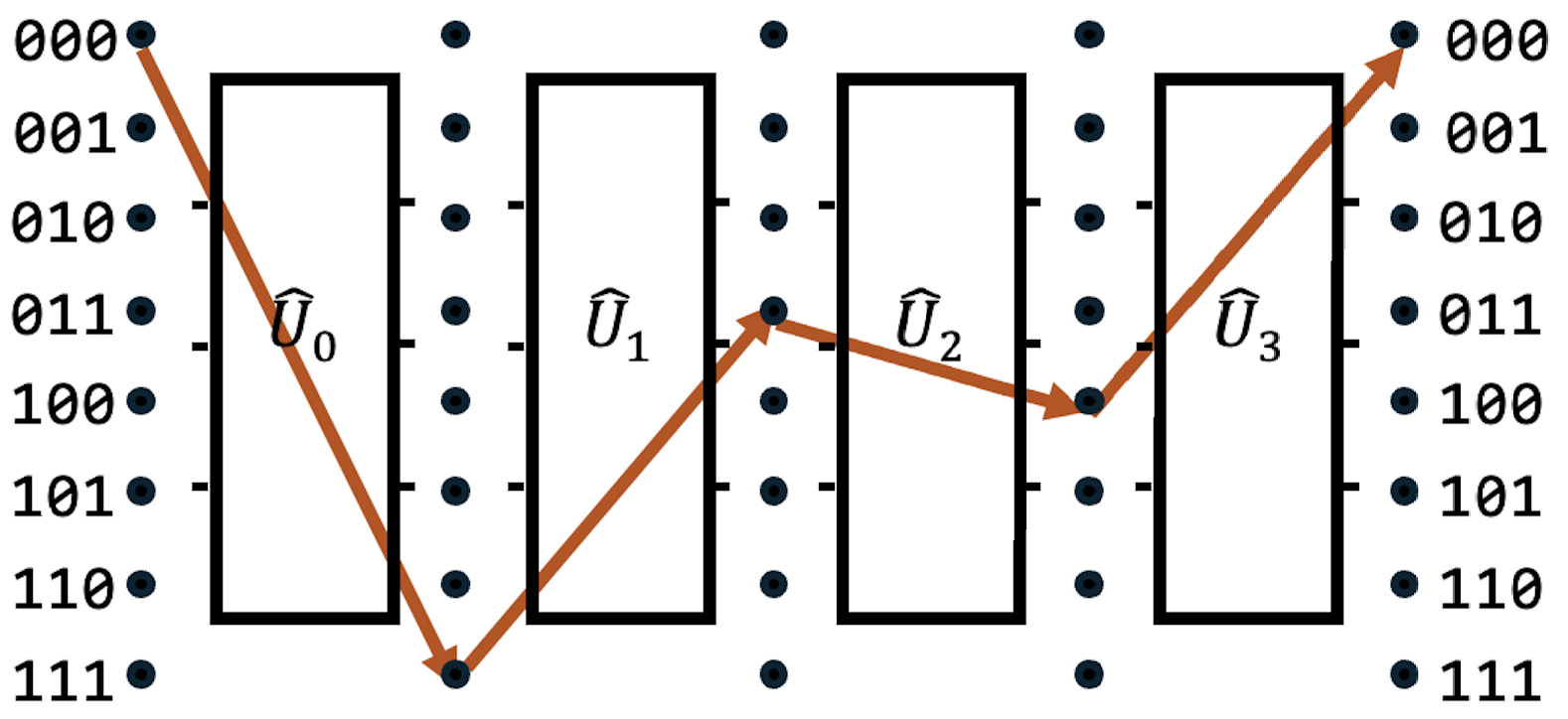}
\caption{A path from $|000\rangle$ to $|000\rangle$ across a circuit with depth $L=4$).}
    \label{fig:PathCircuitExample}
\end{figure}

Figure \ref{fig:PathCircuitExample} depicts an example of a quantum circuit with 3 qubits and 4 layers. It also includes a path connecting the initial computational state $0$ to the final state $0$, through intermediate states $(7, 3, 4)$.

\begin{figure}[htb]
    \centering
    \includegraphics[width=0.9\columnwidth]{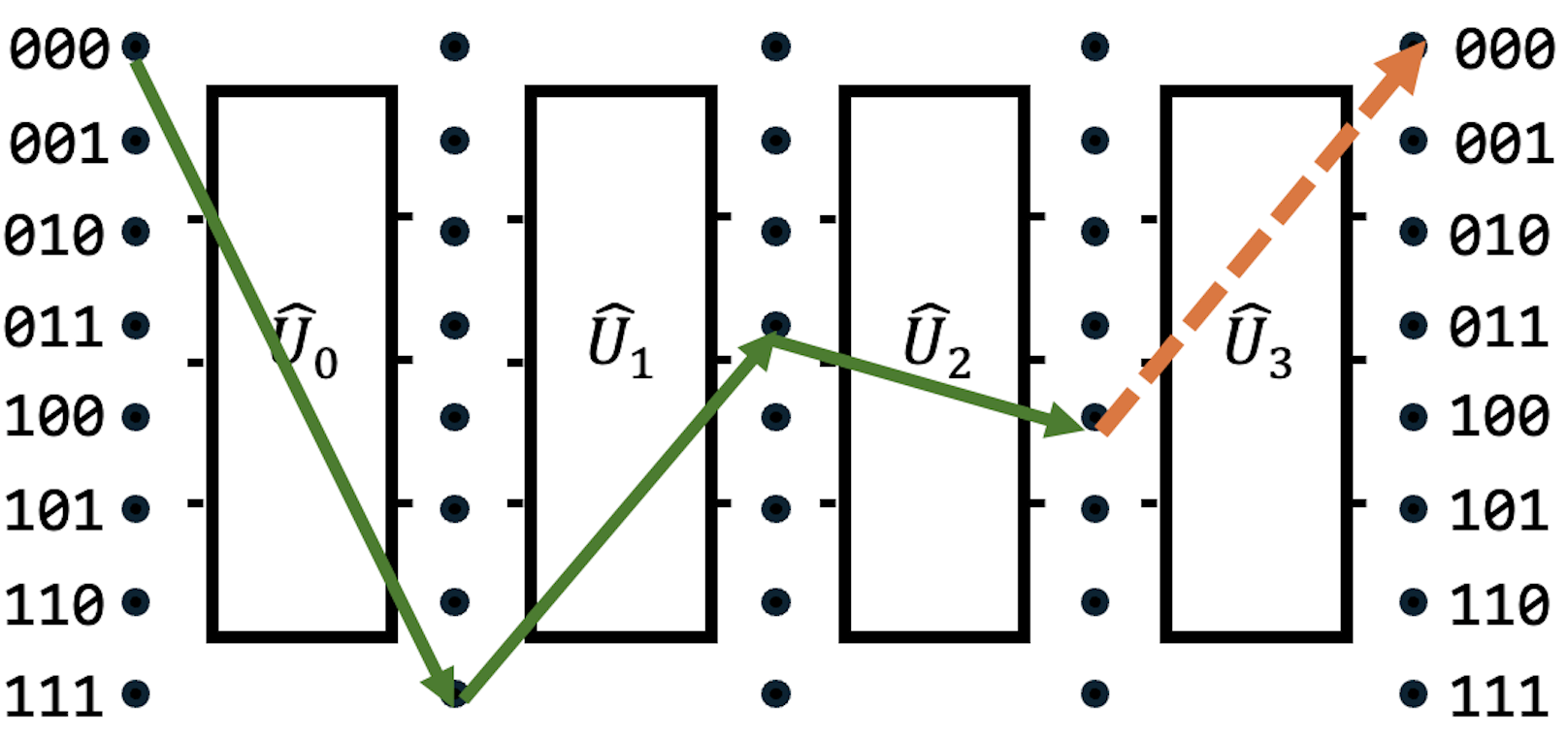}
    \caption{Forward importance sampling: up to layer $L-1$ (green arrows) the path is guaranteed to have both non-zero amplitude and probability. The deterministic connection to the final state (orange dashed arrow) has probability $1$, but its amplitude $\langle \Psi_f | \hat{U}_{L-1}| s_{L-1} \rangle$ can be zero.}
    \label{fig:ForwardIS}
\end{figure}

Paths can be sampled in many different ways. A relatively simple, yet surprisingly effective, approach is to use forward importance sampling. A path is stochastically generated across $L-1$ layers, from the initial state $|\Psi_0 \rangle$ to the next-to-last state, $s_{L-1}$. The output classical state $s_l \in \mathcal{S}$ after each intermediate layer $\hat{U}_{l-1}$ is sampled from the probability distribution $p(s_l | s_{l-1}, \hat{U}_{l-1})=|\langle s_{l}| \hat{U}_{l-1}|s_{l-1} \rangle|^2$, with $s_0=\Psi_0$ and $l=1 \ldots L-1$. Since the final state is fixed, the probability of connecting $s_{L-1}$ to $\Psi_f$ is $1$, although its amplitude $\langle \Psi_f | \hat{U}_{L-1}| s_{L-1} \rangle$ might be $0$ -- see figure \ref{fig:ForwardIS}.

A layer has depth $1$, with each gate operating on a subset of 1 or 2 qubits, and each qubit being acted upon, at most, by a single gate. The layer's unitary is given by the tensor product of the gates' unitaries $\hat{U_l}=\bigotimes_j \hat{G}_{l,(j)}$, where $j$ is a tuple identifying the qubits the gate acts upon.
 
The output classical state $s_l$ after layer $\hat{U}_{l-1}$ is sampled on a per gate basis from the probability distribution defined by equation \ref{eq:forwardSampling} -- note that the short notation $p_{v,v'}$ is used to denote $p(v' \, | v, \, \hat{U})$, to improve readability.
\begin{equation}
\begin{split}
    p_{s_{l-1,(j)},\, s_{l,(j)}} & =p(s_{l,(j)} | s_{l-1,(j)}, \hat{G}_{l-1,(j)}) \\ 
    &= |\langle s_{l,(j)}| \hat{G}_{l-1,(j)}|s_{l-1,(j)} \rangle|^2.
\end{split}
    \label{eq:forwardSampling}
\end{equation}
Consider, for example, the Hadamard operator, $\hat{H} = \frac{1}{\sqrt{2}}\begin{bmatrix}
    1 & 1 \\
    1 & -1
\end{bmatrix}$ and input state $s_{l-1}=|1\rangle$, which acts as a column selector in the operator matrix. In this example, that corresponds to the rightmost column, and the amplitudes of the output state are given by $\frac{1}{\sqrt{2}}[1, \, -1]^T$. The output classical state $s_l$ is sampled from $\{|0\rangle, |1\rangle\}$ using the probability distribution obtained from squaring the magnitudes of the elements of the selected column: 
\[p_{|1\rangle, \, s_l} = p(s_l \, | \, |1\rangle, \hat{H}) =[|\frac{1}{\sqrt{2}}|^2 , \, |-\frac{1}{\sqrt{2}}|^2] = [\frac{1}{2}, \, \frac{1}{2}].\]
Layer $\hat{U}_{l-1}$'s joint sampling distribution is given by 
\[p_{s_{l-1},\,s_l} = \prod_j p(s_{l,(j)} | s_{l-1,(j)}, \hat{G}_{l-1,(j)}).\]

A path sampling probability $p(\bar{z})$ and amplitude $f(\bar{z})$ are given by the respective products across all layers (except the last for $p(\bar{z})$ since $p_{s_{L-1},\Psi_f}=1$):

\begin{equation}
    p(\bar{z}) = \prod_{l=1}^{L-1} p(s_l | s_{l-1}, \hat{U}_{l-1}) \   \label{eq:ISPT_p}
\end{equation}
\begin{equation}
    f(\bar{z}) = \prod_{l=1}^{L} \left| \langle s_l | \hat{U}_{l-1} | s_{l-1} \rangle \right|^2, \, s_0=\Psi_0, \, s_L=\Psi_f.
    \label{eq:ISPT_f}
\end{equation}
The Forward Importance Sampling Monte Carlo (FS) estimator is given by
\begin{equation}
    \langle \Psi_f| \, \hat{U} |\Psi_0 \rangle \approx \tilde{I}_{FS} = \frac{1}{N} \sum_{i=0}^{N-1}    \frac{f\left(\bar{z}_i\right)}{p(\bar{z}_i)}.
\end{equation}

Algorithm \ref{alg:ForwardImportance} details Forward Importance Sampling in pseudo-code.

\begin{algorithm}[htb] 
\SetAlgoLined 
\KwData{U[.], $L$, $n$, $\Psi_0$, $\Psi_f$, $N$} 
\KwResult{Amplitude ($f$)}
$s_0 \leftarrow \Psi_0$ ; $s_L \leftarrow \Psi_f$ \;
$f \leftarrow$ $0. + i0.$ ; 
$s\leftarrow 0$ \;
\While(\tcp*[f]{sample each path $\bar{z}$}){($s < N$)} {
$p(\bar{z}) \leftarrow 1.$ ; $f(\bar{z}) \leftarrow 1.+i0$ \;
isZero $\leftarrow$ false \;
$s_{c} \leftarrow s_0$ \tcp*{current state}  
\For{(l $\leftarrow 0 \dots L-1$) and (not isZero)} {
$p(s_{n}) \leftarrow 1$ \tcp*{$s_{n}$ next state} 
\If(\tcp*[f]{sample next state}){l<L-1}{
$s_{l+1}, p(s_{n}) \leftarrow$ SampleU (U[.], $l$, $s_{c}$);}

$s_{c} \leftarrow s_l$;   
$s_{n} \leftarrow s_{l+1}$\;  
$f_l \leftarrow$ EvalU (U[.], $l$, $s_c$, $s_n$) \;
\eIf(\tcp*[f]{early termination}){$f_l$==0}{isZero $\leftarrow$ true; }{$p(\bar{z}) \leftarrow p(\bar{z}) * p(s_n)$\;
$f(\bar{z}) \leftarrow f(\bar{z}) * f_l$;}

}
\lIf{not isZero}{
$f \leftarrow  f +  \frac{f(\bar{z})}{p(\bar{z})}$}
$s++$ \;
}
$ f\leftarrow \frac{f}{N}$   \tcp*{ average}  
\Return $f$ \tcp*[f]{amplitude}
\caption{Forward Importance Sampling} 
\label{alg:ForwardImportance}
\end{algorithm}

Forward importance sampling convergence rate depends on how likely it is to forward sample a path segment leading to a next-to-last state, $|s_{L-1}\rangle$,  which connects to the final state with non-zero amplitude, i.e., $\langle \Psi_f| \hat{U}_{L-1}|s_{L-1}\rangle \neq 0$.  If such segments are very unlikely, then the non-zero paths have very low probability. When they occur, their contribution to the estimator is large because the amplitude is divided by a small probability $(\ll 1)$, leading to a high-variance Monte Carlo estimate.

The Bidirectional Sampling Monte Carlo estimator proposed next addresses this challenge, minimizing the impact of zero-amplitude deterministic connections to fixed states, such as $\Psi_0$ and $\Psi_f$.

\section{Bidirectional Monte Carlo Path Integration}
\label{sec:BDMC}

Bidirectional Monte Carlo (BMC) algorithms aim to overcome the challenge of hard-to-access final states by exploiting path reversibility. This paper proposes applying BMC techniques to quantum circuit simulation.

A quantum circuit time evolution operator is described by a unitary matrix $\hat{U}$, implying that $\hat{U}^{-1} = \hat{U}^{\dagger}$. Paths can therefore be reversed, implying that transition amplitude estimates can be computed based on paths starting from either $\Psi_0$ or $\Psi_f$, and, in fact, on combinations thereof.

BMC strives to build paths that successfully connect the initial and the final states by combining both forward (from $\Psi_0$) and backward (from $\Psi_f$) sampling. 

\subsection{Bidirectional Importance Sampling}

A bidirectional sample consists of a pair of two path segments, $(\bar{z}_F, \bar{z}_B)$, forward and backward sampled, respectively, with length $L-1$, each. The forward segment is given by \[\bar{z}_F  = (s_0=\Psi_0, s_1, \ldots, s_{L-1}),\] each next state, up to $s_{L-1},$ being sampled according to equation \ref{eq:forwardSampling}. The final deterministic connection from $s_{L-1}$ to $\Psi_f$ across $\hat{U}_{L-1}$ is not included.

The backward segment, labelled using $t$ vertices starting from the final state $t_0 = \Psi_f$, is given by \[\bar{z}_B  = ( t_{L-1}, \ldots, t_1, t_0=\Psi_f),\] each previous state, up to $t_{L-1},$ being sampled according to equation \ref{eq:backwardSampling}, which uses the layers' unitaries adjoints. The final deterministic connection from $t_{L-1}$ to $\Psi_0$ across $\hat{U}_{0}$ is not included.
\begin{equation}
    \begin{split}
     p_{t_{L-l-2},t_{L-l-1}} & = p (t_{L-l-1} | \hat{U}_{l+1}^\dagger, t_{L-l-2}) \\
     & = |\langle t_{L-l-1} | \hat{U}_{l+1}^\dagger| t_{L-l-2} \rangle|^2 ,  
     {\scriptstyle l \in \{L-2 \ldots 0\}}
    \end{split}
    \label{eq:backwardSampling}
\end{equation}

\begin{figure}[htb]
    \centering
\includegraphics[width=0.9\columnwidth]{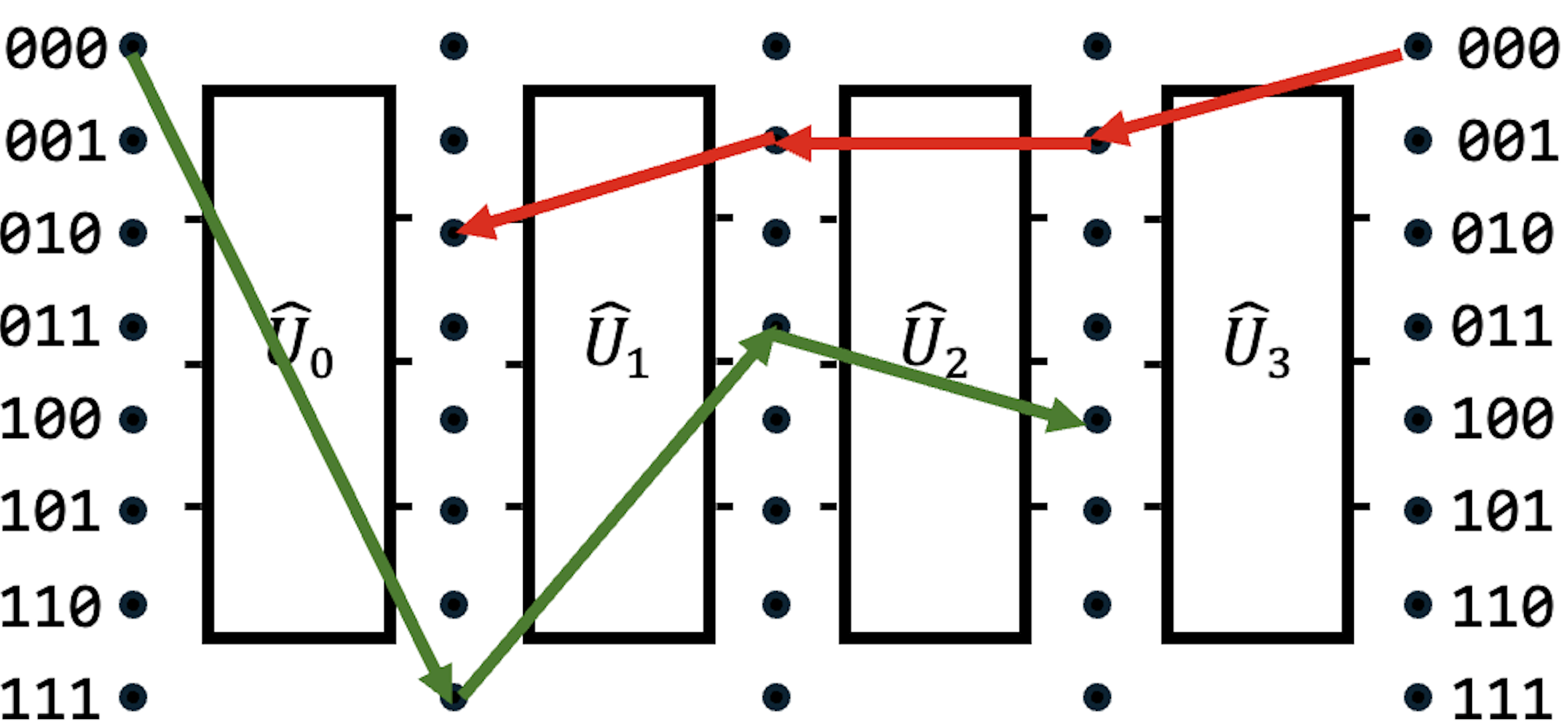}
    \caption{Forward (green) and backward (red) importance sampled path segments for $L=4$.}
    \label{fig:BDPT-FB-paths}
\end{figure}

Figure \ref{fig:BDPT-FB-paths} presents an example of sampled path segments for $L=4$. Each of these path segments is guaranteed to have non-zero amplitude and probability.

\begin{figure}[htb]
    \centering
\begin{subfigure}{0.493\columnwidth}
    \includegraphics[width=\textwidth]{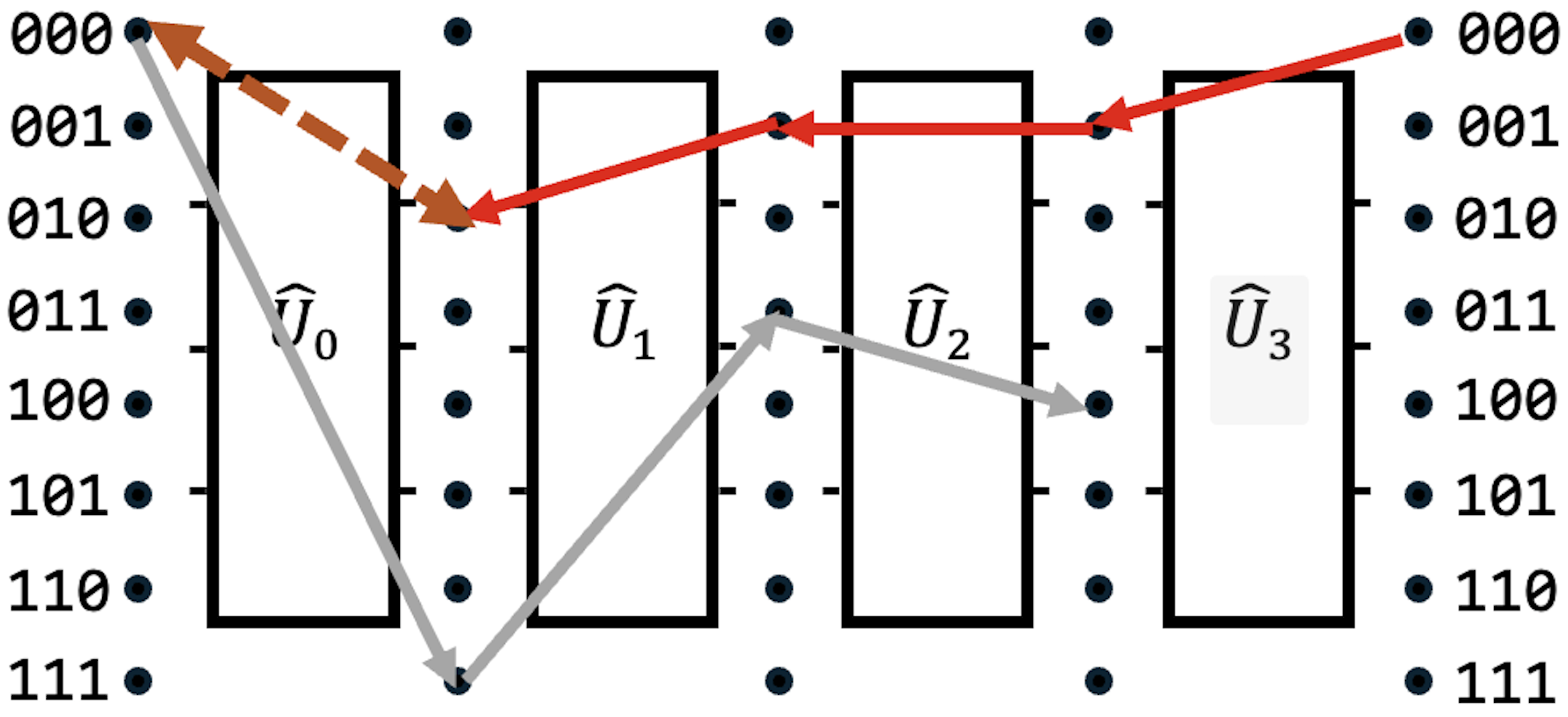}
    \caption{Connect across L0}
    \end{subfigure}
\hfill
\begin{subfigure}{0.493\columnwidth}
    \includegraphics[width=\textwidth]{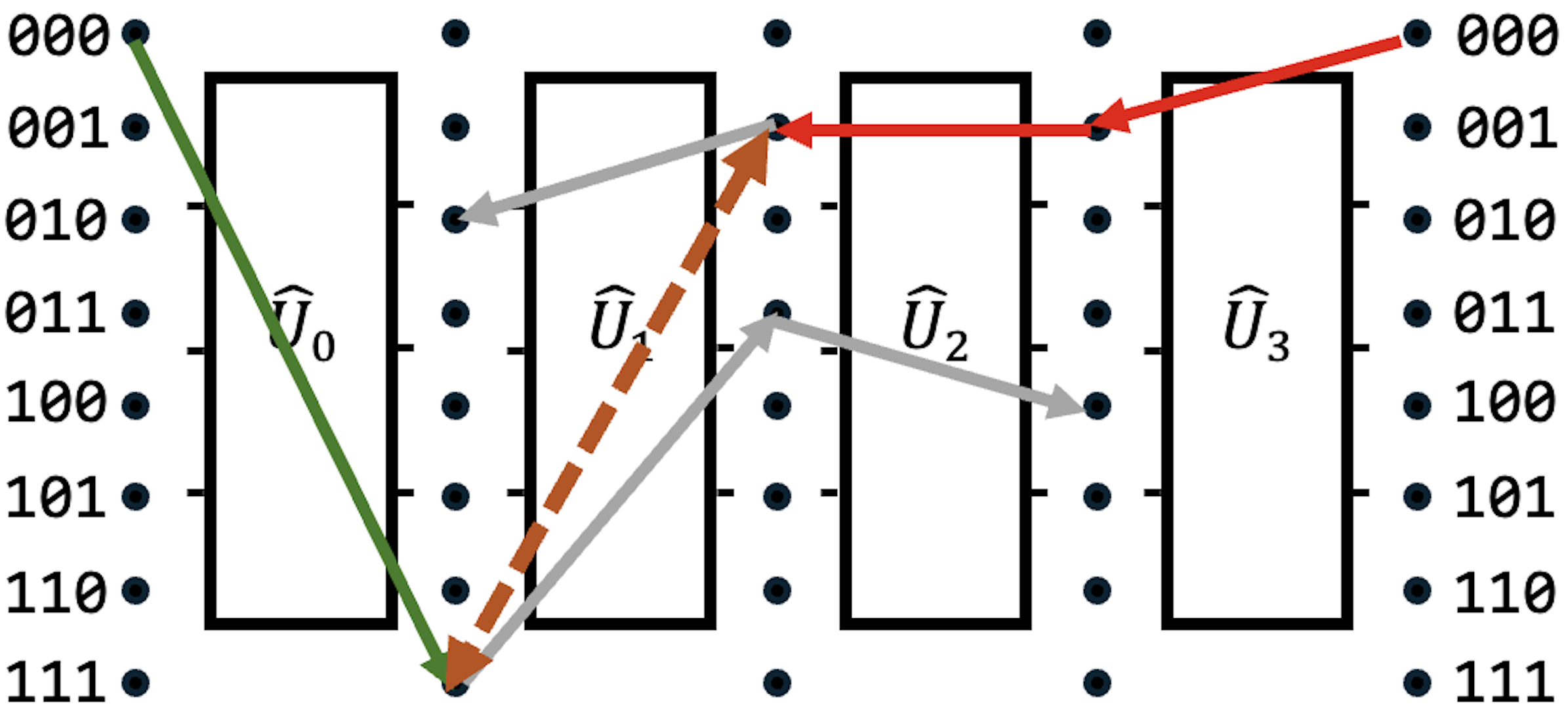}
    \caption{Connect across L1}
    \label{subfig:BDPT-L1}
\end{subfigure}
\\
\begin{subfigure}{0.493\columnwidth}
    \includegraphics[width=\textwidth]{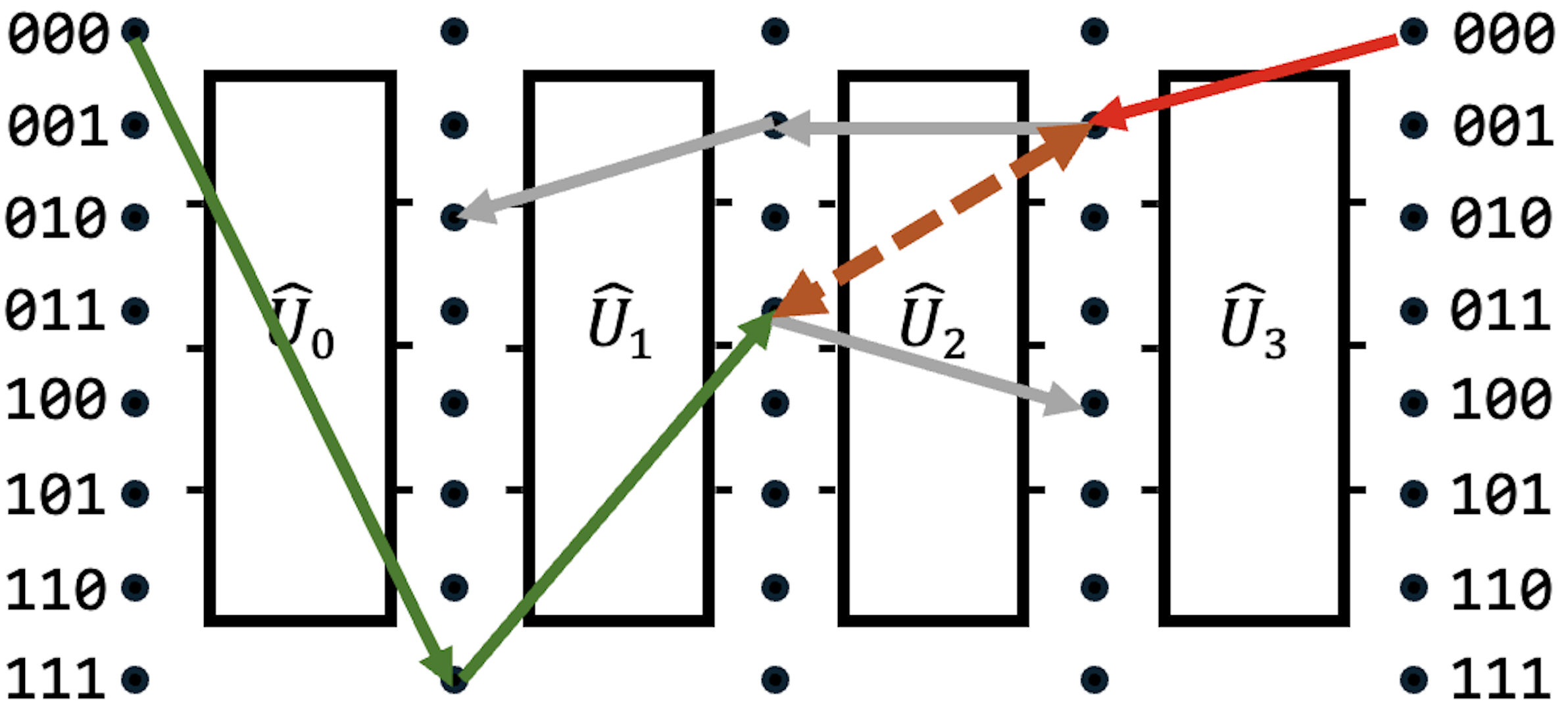}
    \caption{Connect across L2}
\end{subfigure}
\hfill
\begin{subfigure}{0.493\columnwidth}
    \includegraphics[width=\textwidth]{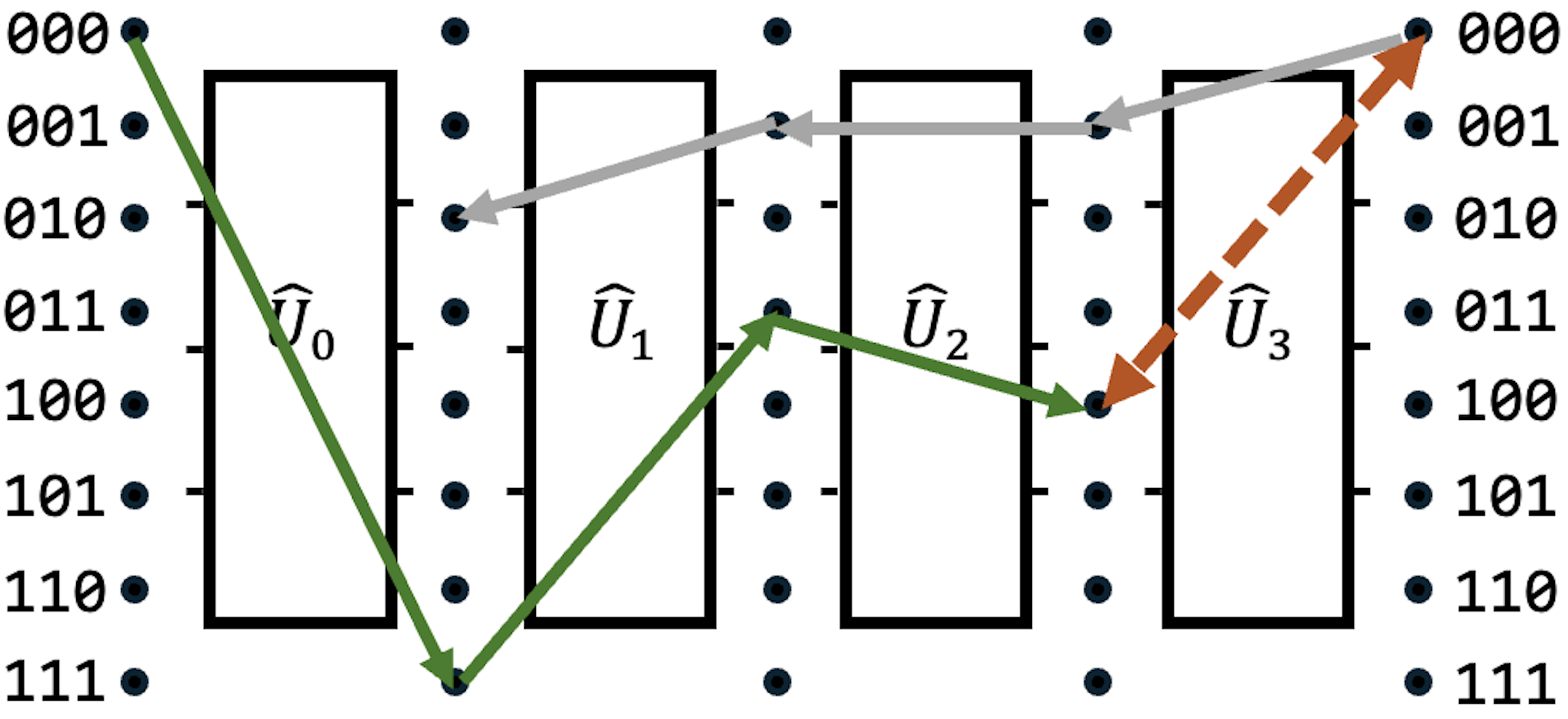}
    \caption{Connect across L3}
\end{subfigure}
    \caption{Paths generated by connecting forward and backward path segments. The dashed dark orange arrow represents the deterministic connection. Non-used path segments are dimmed to gray.}
    \label{fig:BDPT_Lconnect}
\end{figure}

$L$ full paths are built by deterministically connecting $\bar{z}_F$'s state $s_l$ to $\bar{z}_B$'s state $t_{L-l-1}$, with $l=0 \ldots L-1$ and $p_{s_l, t_{L-l-1}}=1$ since the connection is deterministic. The sampling probability of such path $\bar{z}_{l}$ is given by
\begin{equation}
        p(\bar{z}_l)   = \prod_{j=0}^{l-1} \bigr(p_{s_j, s_{j+1}}\bigl) \, * \prod_{j=l+1}^{L-1} \bigr(p_{t_{L-j}, t_{L-j-1}}\bigl) , {\scriptstyle l=0 \ldots L-1}
    \label{eq:BD-probs}
\end{equation}
and its amplitude is
\begin{equation}
\begin{split}    
        f(\bar{z}_l)  = & \prod_{j=0}^{l-1} \bigr( \langle s_{j+1} | \hat{U}_j | s_j \rangle \bigl) \, * 
         \langle t_{L-l-1} | \hat{U}_l | s_l \rangle \, * \\
        & \prod_{j=l+1}^{L-1} \bigr(\langle t_{L-j} | \hat{U}_j^\dagger | t_{L-j-1} \rangle \bigl) , {\scriptstyle l=0 \ldots L-1}
    \label{eq:BD-amps}
\end{split}
\end{equation}
Figure \ref{fig:BDPT_Lconnect} illustrates the 4 paths built from the example in figure \ref{fig:BDPT-FB-paths}.

The potential advantage of Bidirectional Importance Sampling (BIS) over FS stems from the fact that each of these $L$ paths is more likely to have a non-zero contribution than forward only paths, since it combines bidirectional data.  Veach shows that the BIS estimator, given by equation \ref{eq:QBDPT_estimator}, is unbiased \cite[Chapter 9]{Veach1997}. 
\begin{equation}
\tilde{I}_{BIS} = \frac{1}{N*L} \sum_{i=0}^{N-1} \sum_{l=0}^{L-1}    \frac{f\left(\bar{z}_{i,l}\right)}{p(\bar{z}_{i,l})}.
\label{eq:QBDPT_estimator}
\end{equation}

Algorithm \ref{alg:BDPT} depicts this approach in pseudo-code.

\begin{algorithm}[htb] 
\SetAlgoLined 
\KwData{U[.], $L$, $n$, $\Psi_0$, $\Psi_f$, $N$} 
\KwResult{Amplitude ($f$)}
f $\leftarrow$ $0. + i0.$ ; 
$s \leftarrow 0$ \;
\While{($s < N$)} {
\tcp{$pathF$ -- forward $\bar{z}_F = (\psi_0, \ldots s_{L-1})$}
\tcp{$pF[l]$ -- probability from $\psi_0$ to $s_l$}
\tcp{$aF[l]$ -- amplitude from $\psi_0$ to $s_l$}
$pathF, pF, aF \leftarrow$ GPathF (U[.], $L$, $n$, $\Psi_0$) \;  
\tcp{$pathB$ {\small -- backward} $\bar{z}_B = (t_{L-1}, \ldots, \Psi_f)$}
\tcp{$pB[{\scriptstyle L-l-1}]$ -- {\small probability} $\Psi_f \rightarrow t_{\scriptstyle L-l-1}$}
\tcp{$aB[{\scriptstyle L-l-1}]$ -- {\small amplitude} $\Psi_f \rightarrow t_{\scriptstyle L-l-1}$}
$pathB, pB, aB \leftarrow$ GPathB (U[.], $L$, $n$, $\Psi_f$) \; 

\For{(l $\leftarrow 0 \dots L-1$)} {
\tcp{connect across layer $l$}
$aC \leftarrow$ Connect (U[.], $l$, $pathF$, $pathB$) \;
$f(\bar{z}) \leftarrow aF[l] * aC * aB[l]$ \;
$p(\bar{z}) \leftarrow pF[l] * 1.0 * pB[l] $\;

$ f \leftarrow f + \frac{f(\bar{z})}{p(\bar{z})}$   \tcp*{accummulate} 
}
$s++$ \;
}
$f \leftarrow \frac{f}{N * L}$   \tcp*{average}  
\Return $f$ \tcp*[f]{amplitude}
\caption{Bidirectional Importance Sampling} 
\label{alg:BDPT}
\end{algorithm}

\subsection{Multiple Importance Sampling}

The previous section generated complete paths by progressively connecting two path segments. While this is effective, variance can be further reduced via the use of Multiple Importance Sampling (MIS) \cite{Veach1997}. This is a powerful variance reduction technique 
that further improves the $\tilde{I}_{BIS}$ estimator by considering all the ways the path could have been generated. 

Paths sampled with low probability are the major sources of variance for the previously described Monte Carlo estimators. If a connection within a path is sampled with very low probability, then the whole path sampling probability is very low, $p(\bar{z}) \approx 0$. Vanilla Monte Carlo estimators evaluate $\bar{z}$'s contribution as
$\tilde{I}_{\bar{z}} = \frac{f(\bar{z})}{p(\bar{z})}$, which overestimates the average value, $\tilde{I}_{\bar{z}} \gg \tilde{I}$, resulting in large variance, thus requiring a large number of samples, $N$, to converge. 

Each bidirectional path is built by deterministically connecting a forward path segment with a backward path segment across a specific layer $\hat{U}_l$. MIS exploits the fact that the same path could have been generated by deterministically connecting across any other layer, which would result in $L$ different sampling strategies, by moving the deterministic connection through the layers. The same path is therefore associated with $L$ different sampling probabilities, $p_d(\bar{z}) \, , \, {\scriptstyle d = 0 \ldots L-1}$, each resulting from deterministically connecting $s_d$ to $t_{L-d-1}$. $d$ indexes the layer $\hat{U}_d$ associated with the hypothetical deterministic connection. Each probability $p_d(\bar{z}_l)$ is given by:
\begin{equation}
        p_d(\bar{z}_l)   = \prod_{v=0}^{d-1} \bigr(p_{q_{v+1}, q_{v}}\bigl) \, * \prod_{v=d+1}^{L-1} \bigr(p_{q_v, q_{v+1}}\bigl) , {\scriptstyle d=0 \ldots L-1}
        \label{eq:MIS_pd}
\end{equation}
with $\{q_0, \ldots , q_L\} = \{s_0,\ldots, s_l, t_{L-l-1},\ldots , t_0\}$. Note that the path amplitude $f(\bar{z}_l)$ is independent of $d$, since only sampling probabilities change.

\begin{figure}[htb]
    \centering
\begin{subfigure}{0.49\columnwidth}
    \includegraphics[width=\textwidth]{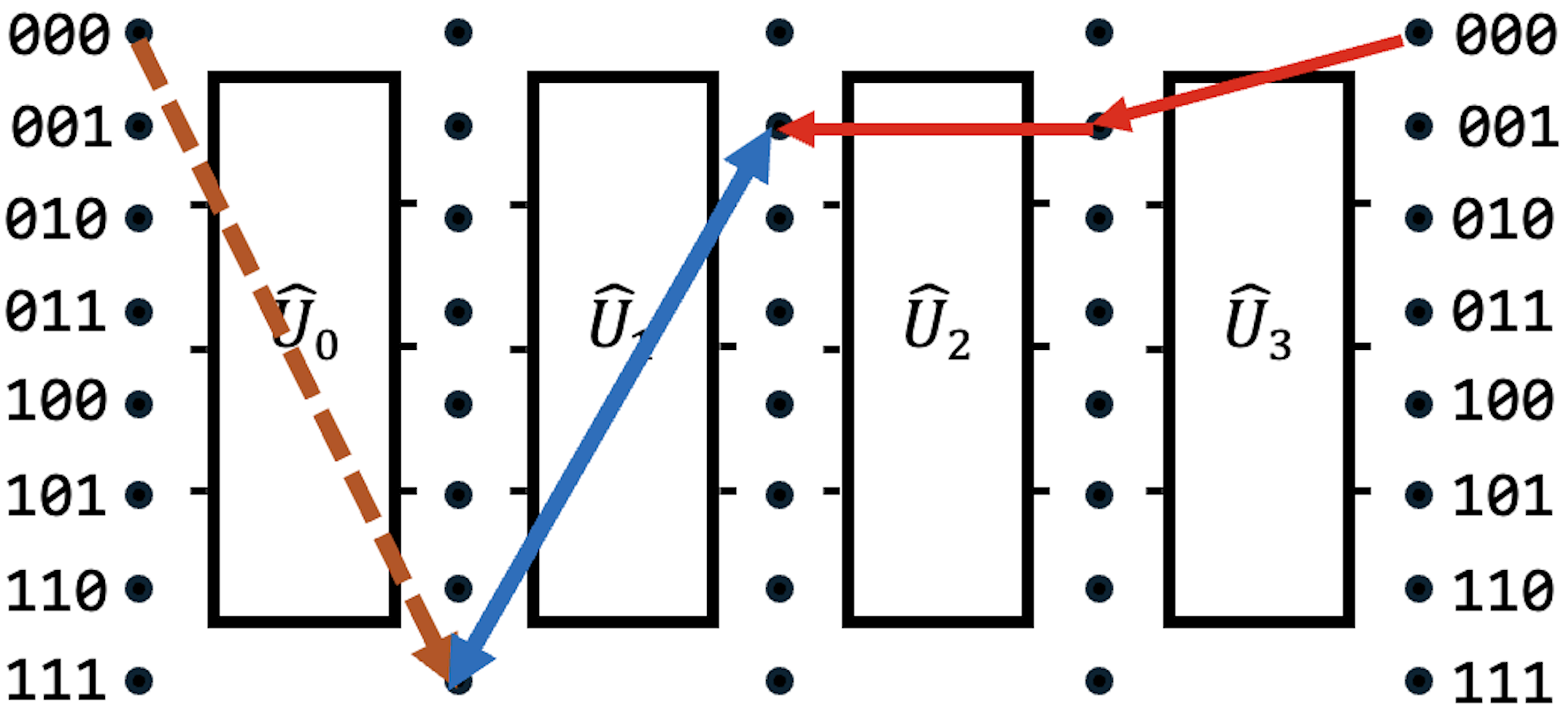}
    \caption{Connected across L0}
    \label{subfig:MIS_L0}
    \end{subfigure}
\hfill
\begin{subfigure}{0.49\columnwidth}
    \includegraphics[width=\textwidth]{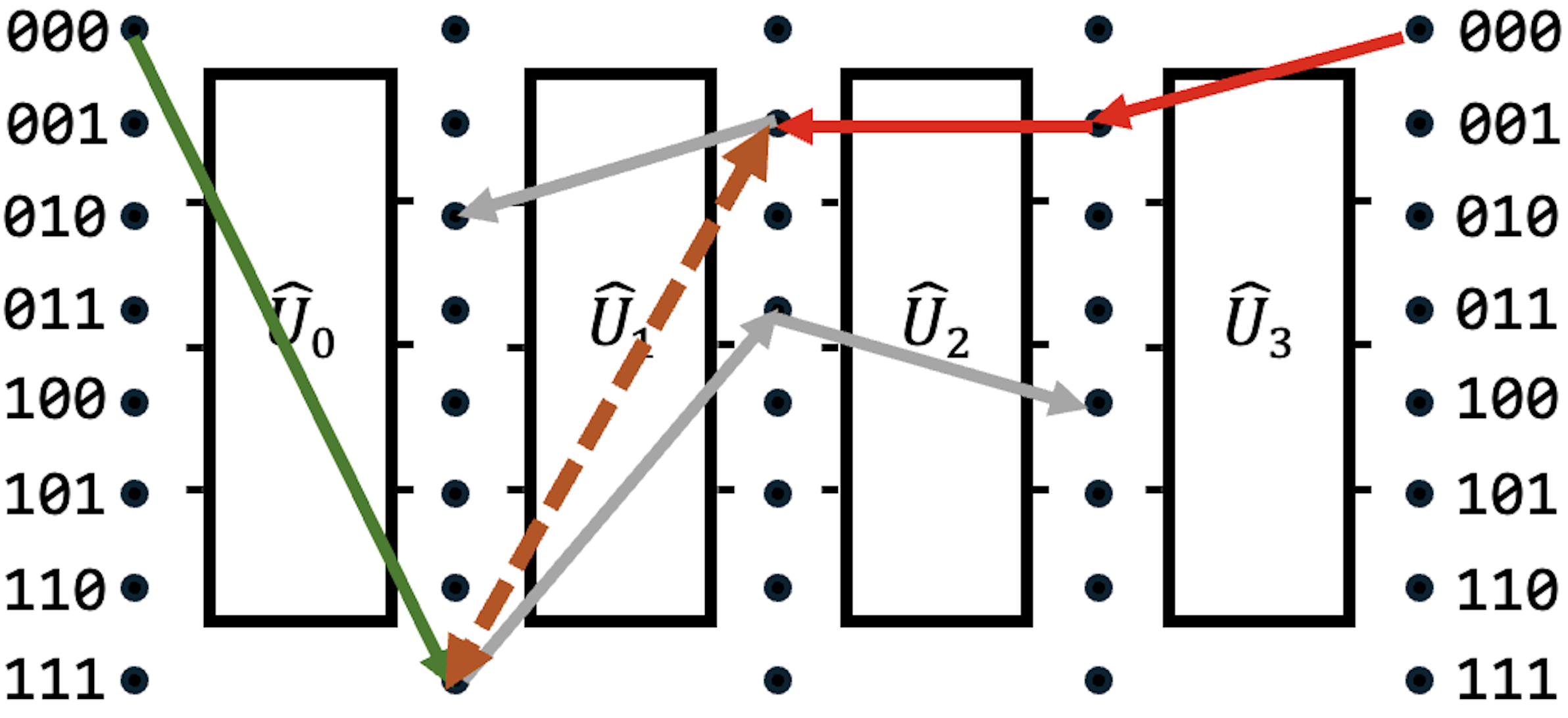}
    \caption{Connected across L1}
    \label{subfig:MIS_L1}
\end{subfigure}
\\
\begin{subfigure}{0.49\columnwidth}
    \includegraphics[width=\textwidth]{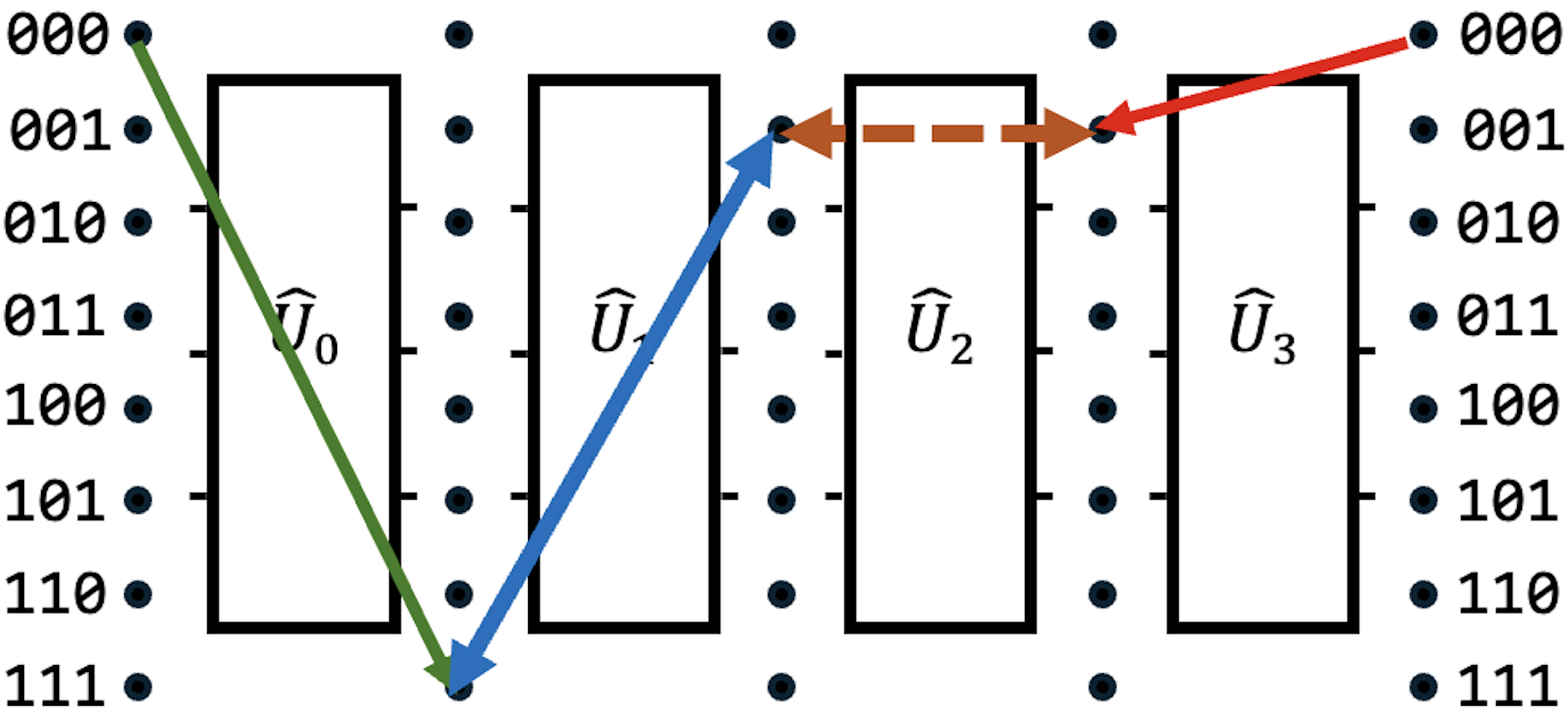}
    \caption{Connected across L2}
    \label{subfig:MIS_L2}
\end{subfigure}
\hfill
\begin{subfigure}{0.49\columnwidth}
    \includegraphics[width=\textwidth]{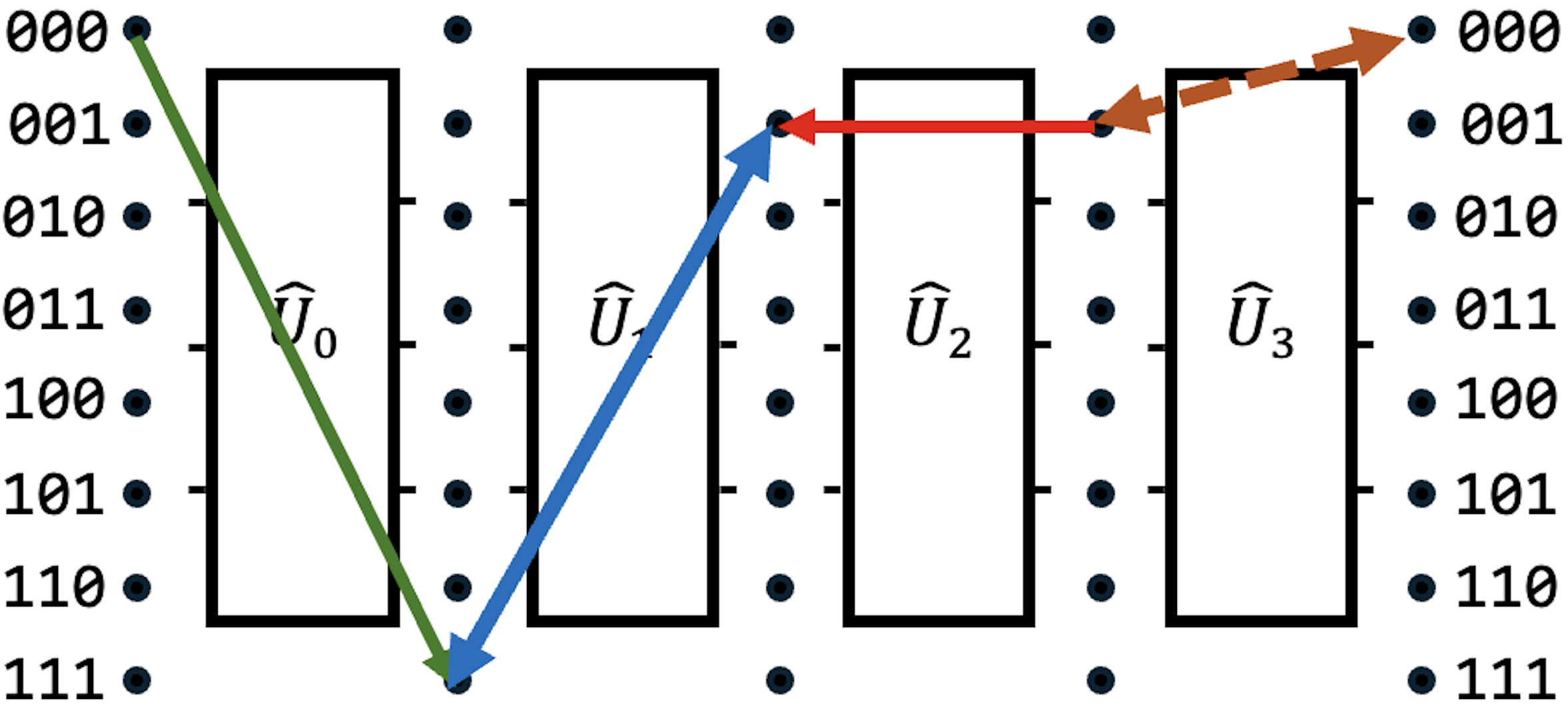}
    \caption{Connected across L3}
    \label{subfig:MIS_L3}
\end{subfigure} 
    \caption{$L$ different manners the same path, originally deterministically connected across L1 as depicted by the blue arrow (see figure \ref{subfig:BDPT-L1}), could have been generated by shifting the deterministic connection (dark orange dashed arrow).}
    \label{fig:BDPT_MIS_paths}
\end{figure}

Figure \ref{fig:BDPT_MIS_paths} illustrates, for the path depicted in figure \ref{subfig:BDPT-L1} and effectively created by a deterministic connection across $\hat{U}_l, \, l=1$, the 4 different manners in which it could have been generated. The respective probabilities (equation \ref{eq:MIS_pd}) are:
\begin{align*}
    p_{d=0}(\bar{z}) &= 1 * p_{t_2, s_1} * p_{t_2, t_1} * p_{t_1, t_0} \\
    p_{d=1}(\bar{z}) &= p_{s_1, s_0} * 1 *  p_{t_2, t_1} * p_{t_1, t_0} \\
    p_{d=2}(\bar{z}) &= p_{s_1, s_0} * p_{t_2, s_1} * 1 * p_{t_1, t_0} \\
    p_{d=3}(\bar{z}) &= p_{s_1, s_0} * p_{t_2, s_1} *  p_{t_2, t_1} * 1
\end{align*}

The MIS estimator combines the $L$ different sampling probabilities, $p_d(\bar{z})$, for each sampled path $\bar{z}$, to allow for a robust estimate.  Equation \ref{eq:renderingMCpaths2}
 adds an weight term $w(\bar{z}_i)$ to the MC estimator introduced with equation \ref{eq:MCestimator}:
 
 \begin{equation}
    \langle \Psi_f | \hat{U} | \Psi_0 \rangle \approx \tilde I_{MC} = \frac{1}{N}\sum_{i=0}^{N-1} w(\bar{z}_i) \frac{f(\bar{z}_i)}{p(\bar{z}_i)},
    \label{eq:renderingMCpaths2}
\end{equation}
with $p(\bar{z}_i)$ standing for the probability with which $\bar{z}_i$ was actually sampled. The balance heuristic \cite{Veach1997} proposes weights $w(\bar{z}_i)$ that make the sample contribution independent of $p(\bar{z}_i)$:
\begin{equation}
    w(\bar{z}_i) = \frac{p(\bar{z}_i)}{\sum_{d=0}^{L-1} p_d (\bar{z}_i)}
    \label{eq:balance_heuristic}
\end{equation}

By substituting \ref{eq:balance_heuristic} into equation \ref{eq:renderingMCpaths2} we get the   Multiple Importance Sampling (MIS) estimator
\begin{equation}
\begin{split}
    \tilde{I}_{MIS} & = \frac{1}{N} \sum_{i=0}^{N-1} w(\bar{z}_i) \frac{f(\bar{z}_i)}{p(\bar{z}_i)} \\
    & = \frac{1}{N} \sum_{i=0}^{N-1} \frac{p(\bar{z}_i)}{\sum_{d=0}^{L-1} p_d (\bar{z}_i)} \frac{f(\bar{z}_i)}{p(\bar{z}_i)} \\
    & = \frac{1}{N} \sum_{i=0}^{N-1} \frac{f(\bar{z}_i)}{\sum_{d=0}^{L-1} p_d (\bar{z}_i)}
\end{split}
    \label{eq:MIS_estimator}
\end{equation}

The denominator $\sum_{d=0}^{L-1} p_d (\bar{z}_i)$ in equation \ref{eq:MIS_estimator} is always larger than or equal to the denominator of the non MIS estimator (equation \ref{eq:renderingMCpaths2} with $w(\bar{z}_i) = 1$), therefore $\tilde{I}_{MIS}$ is numerically more stable, exhibiting a better variance convergence rate. 

Veach and Guibas \cite{Veach1995} show that the balance heuristic is near-optimal in the sense that its variance is bounded relative to that of other valid MIS estimators.  

Applying MIS to the BMC simulation of quantum circuits results in the final Bidirectional Sampling estimator (BS):
\begin{equation}
\tilde{I}_{BS} = \frac{1}{N*L} \sum_{i=0}^{N-1} \sum_{l=0}^{L-1}    \frac{f\left(\bar{z}_{i,l}\right)}{\sum_{d=0}^{L-1}p_d(\bar{z}_{i,l})}
\label{eq:BS_estimator}
\end{equation}

Algorithm \ref{alg:BS} illustrates the approach in pseudo-code. Each of the $L$ paths built by bidirectional sampling (line \ref{algline:LPaths}) requires the sum of $L$ different path generation probabilities (line \ref{algline:MIS_ForLoop}).

\begin{algorithm}[htb] 
\SetAlgoLined 
\KwData{U[.], $L$, $n$, $\Psi_0$, $\Psi_f$, $N$} 
\KwResult{Amplitude ($f$)}
$s_0 \leftarrow \Psi_0$ ;  $s_L \leftarrow \Psi_f$ \;
$f \leftarrow 0. + i0.$ ; 
$s \leftarrow 0$ \;
\While{($s < N$)} {
\tcp{$pF[l]$ -- probability from $\psi_0$ to $s_l$}
\tcp{$aF[l]$ -- amplitude from $\psi_0$ to $s_l$}
$pathF, pF, aF \leftarrow$ GPathF (U[.], $L$, $n$, $\Psi_0$) \;  
\tcp{$pB[{\scriptstyle L-l-1}]$ -- {\small probability} $\Psi_f \rightarrow t_{\scriptstyle L-l-1}$}
\tcp{$aB[{\scriptstyle L-l-1}]$ -- {\small amplitude} $\Psi_f \rightarrow t_{\scriptstyle L-l-1}$}
$pathB, pB, aB \leftarrow$ GPathB (U[.], $L$, $n$, $\Psi_f$) \; 

\For{(l $\leftarrow 0 \dots L-1$)} { \tcp{connect across layer $l$}
\label{algline:LPaths}
$aC \leftarrow$ Connect (U[.], $l$, $pathF$, $pathB$) \;
$f(\bar{z}) \leftarrow aF[l] * aC * aB[l]$ \;

sumProb $\leftarrow$ 0 \;
\For{(d $\leftarrow 0 \dots L-1$)} {
\label{algline:MIS_ForLoop}
$p(\bar{z}) \leftarrow pF[d] * 1.0 * pB[d] $\;
sumProb $\leftarrow$ sumProb + $p(\bar{z})$ \;
}

$f \leftarrow f + \frac{f(\bar{z})}{\mbox{sumProb}}$   \tcp*{accumulate} 
}
$s++$ \;
}
$f \leftarrow \frac{f}{N * L}$   \tcp*{average}  
\Return $f$ \tcp*{amplitude}
\caption{Bidirectional Multiple Importance Sampling (BS)} 
\label{alg:BS}
\end{algorithm}

MIS plays a fundamental role in increasing BMC convergence rate. Using a sum of probabilities in the estimator's denominator steers the computation away from divisions by small probabilities, which would otherwise yield large ratios and therefore introduce peaks in the variance. It is a robust, unbiased estimator often used by the computer graphics community to render physically based high quality images. We claim that it can be equally applied to the simulation of quantum circuits.

\section{Experimental Methodology}

This section describes the experimental methodology used to corroborate or falsify the three hypotheses formulated in section \ref{sec:Introduction}. The Monte Carlo simulators were written in C++, using gcc 12.3.0 together with OpenMP to exploit multithreading parallelism and MPI to distribute the workload across multiple computing nodes. Jobs were executed on the \href{ https://deucalion.acnca.pt/pt/}{Deucalion supercomputer}, using up to 16 Intel x86 nodes, with 128 physical processing cores each.

\subsection{H1 -- Convergence rate} 
\label{subsec:MethodH1}

{\em $\tilde{I}_{BS}$ achieves better convergence than $\tilde{I}_{FS}$ when estimating transition amplitudes for quantum circuits.}
\vspace{3mm}

To corroborate or falsify this hypothesis, experimental results are obtained by running both estimators on a range of circuits. 

Monte Carlo's convergence is measured by the variance of the estimator, given by 
\begin{equation}
    Var_{\estimate{I}} = \frac{1}{E} \sum_{e=0}^{E-1} \left( I - \estimate{I}_e \right)^2
\end{equation}
where $E$ is the number of experiments, $I$ is the true value of the integral and $\estimate{I}_e$ is the estimate of the $e^{th}$ experiment. We use $E=5$.

The two algorithms' convergence rates are compared using two different metrics: variance as a function of the number of samples $N$, and variance as a function of execution time $T$. 

Data is collected using two different families of circuits, representative of different features and challenges concerning quantum state evolution: 
\begin{description}
    \item[Instantaneous Quantum Polynomial]  (\IQP) {\bf --}  kernel inversion test for two vectors encoded using the IQP kernel. The overlap is estimated by the probability of measuring $|0\rangle$. IQP circuits are most likely hard to simulate classically \cite{Bremner2011,Marshall2024}. 
    \item[Random]  (\RND) {\bf --}  generated by randomly sampling gates from the Clifford+T gate set, according to the rules defined in \cite{Boixo2018}. Random circuits are known to be exponentially hard to simulate classically (in the ideal, noise-free case) \cite{Arute2019}.
\end{description}

For each circuit family, three different sizes are used, differing in the number of qubits ($n$) and depth ($L$). Table \ref{tab:circuitsConv} presents detailed data on each circuit. See section \ref{sec:dataavailability} for data availability.
For each circuit, we also report the total number of paths connecting $|\Psi_0\rangle$ to $|\Psi_f\rangle$ and the number of such paths that actually have non-zero amplitude (\textbf{\#NZ}), allowing quantification of path space sparsity. Evaluating \textbf{\#NZ} is a computationally very demanding task -- values are not available (n.a.) for larger circuits.

\begin{table}[htb]
    \centering
    \begin{tabular}{cc}
        \textbf{(ID, $\mathbf{n,L}$)}  & \textbf{\#paths} / \textbf{\#NZ} \\ \hline
  
        {\small (\textbf{IQP5},5,17)} & {\small $1E24 /\, 32$}\\ 
         {\small (\textbf{IQP15},15,55)} & {\small $7E243 / \approx 3E4$}  \\ 
         {\small (\textbf{IQP25},25,95)} & {\small $1E830$ / n.a.}  \\ \hline
        {\small (\textbf{RND8},8,9)} & {\small $2E19 / \, 512$}  \\ 
         {\small (\textbf{RND10},10,13)} & {\small $1E43 / \approx 2E5$} \\ 
         {\small (\textbf{RND12},12,15)} & {\small $1E61 /\approx 1E9$}  \\ \hline           \end{tabular}
    \caption{Circuits for experimental assessment of the convergence rate {\small ($\mathbf{n}$ - qubits, $\mathbf{L}$ - layers, {\bf \#paths} - total number of paths, {\bf \#NZ} -- number of non-zero paths)}. }
    \label{tab:circuitsConv}
\end{table}

\subsection{H2 -- Simulation of large circuits} 

{\em $\tilde{I}_{BS}$ allows estimating transition amplitudes of very large quantum circuits (e.g., 4096 qubits) because its memory complexity is $\mathcal{O}(nL)$ and it efficiently finds non-zero paths in a very large and sparse path space.}
\vspace{3mm}

Evaluating $\langle \Psi_f | \hat{U} | \Psi_0 \rangle$ for circuits with hundreds or thousands of qubits is not possible using commonly available classical simulators.  To verify the correctness of the proposed algorithms, experiments are performed using deterministic Hidden Shift Circuits (\HSC), which prepare a classical state, $|\Psi_f \rangle = j, \, {\scriptstyle j \in \{0, \ldots , 2^n-1\}}$ that is known \emph{a priori} \cite{Peres2023}. The simulator's output can thus be verified against the correct, known output. Table \ref{tab:circuitsHSC} presents detailed data on each used Hidden Shift Circuit.

\begin{table}[htb]
    \centering
    \begin{tabular}{cc}
        \textbf{(ID, $\mathbf{n,L}$)}  & \textbf{\#paths}  \\ \hline
  
        {\small (\textbf{HSC256},256,43)} & {\small $1E3600$} \\ 
        {\small (\textbf{HSC1024},1024,43)} & {\small $1E14400$} \\ 
        {\small (\textbf{HSC4096},4096,43)} & {\small $1E57400$} \\ \hline
         \end{tabular}
    \caption{Large Hidden Shift Circuits  {\small ($\mathbf{n}$ - qubits, $\mathbf{L}$ - layers, {\bf \#paths} - total number of paths}. }
    \label{tab:circuitsHSC}
\end{table}

\subsection{H3 -- Sparsity and numerical sign problem} 
\label{subsec:MethodH3}

{\em $\tilde{I}_{BS}$ improvements on the convergence rate, over  $\tilde{I}_{FS}$, are due to a more efficient sampling of the sparse path space, rather than to a better handling of destructive interference, i.e., MC's numerical sign problem.}
\vspace{3mm}

\begin{figure}[htb]
    \includegraphics[width=\columnwidth]{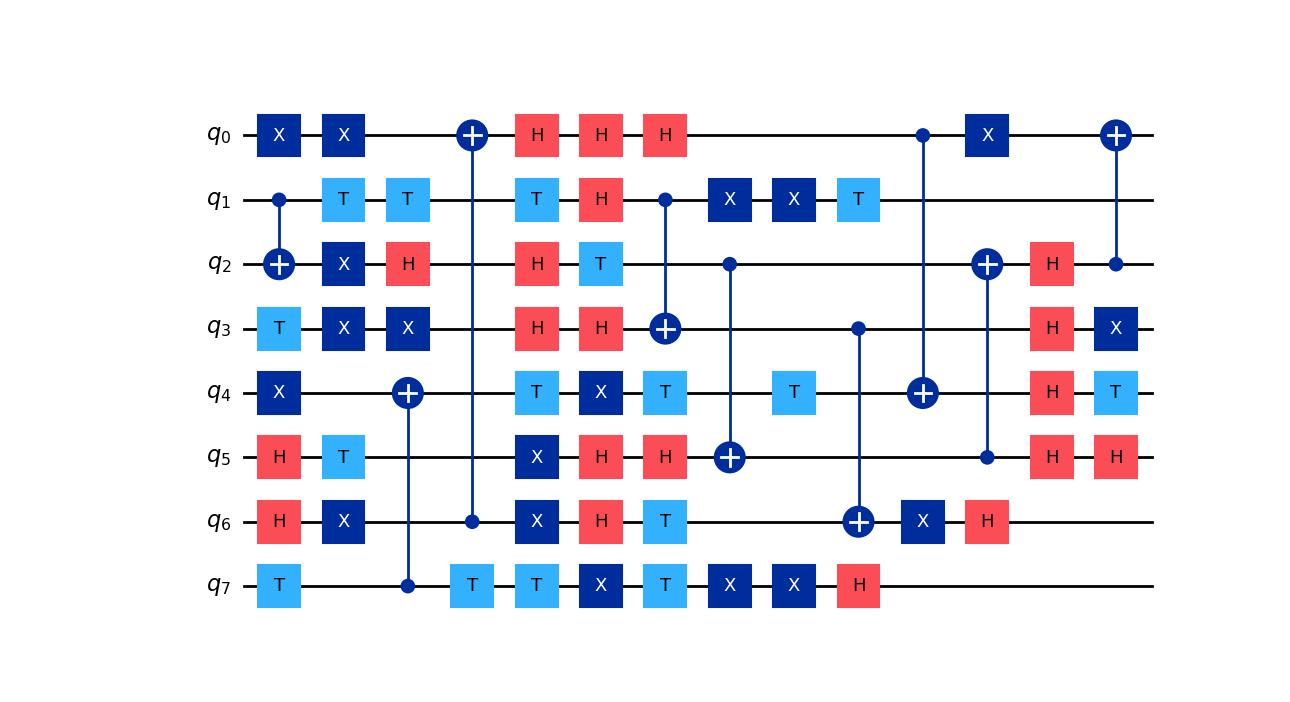}
    \caption{Example random circuit used to assess how sparsity and destructive interference impact MC estimators' convergence.}
    \label{fig:ExampleRndCirc}
\end{figure}

To assess the impact of non-classicality on simulation performance, an additional set of 500 random circuits is used. All circuits have the same shape: 8 qubits and depth 9. Gates are randomly drawn from a pool of $X$, $H$, $T$, and $CX$ gates, with probabilities $[0.28, 0.28, 0.28, 0.16]$, respectively, whilst simultaneously making sure that each qubit is acted upon by exactly 9 gates. 
All circuits share several properties, such as shape and total number of paths associated with any $\langle \Psi_f | \hat{U} | \Psi_0 \rangle$ $(2^{n*(L-1)}=2^{64})$. The different gates' layouts, however, result in wide ranges of sparsity and inverse squared phase, $\chi$ -- see section \ref{subsec:hardness}. The ratio $\frac{Var_{FS}}{Var_{BS}}$ will be used to assess the impact these properties have on the estimators' relative convergence rates.
Figure \ref{fig:ExampleRndCirc} illustrates one such circuit.

\section{Results}
\label{sec:Results}

Empirical data were collected using the two algorithms for the different circuits. Results' analysis is organized into a comparison of the algorithms' convergence, handling of very large circuits, and the impact of the numerical sign problem.

To facilitate visual interpretation data obtained for the $\tilde{I}_{FS}$ estimator is presented in shades of red. In contrast, data for the $\tilde{I}_{BS}$ estimator is presented in shades of green. The saturation of the respective hue represents the size of the circuit associated with each curve: smaller circuits are depicted with less saturated colors.

\subsection{H1 -- Convergence rate} 

\begin{figure}[htb]
    \centering
\begin{subfigure}{\columnwidth}
    \includegraphics[width=\textwidth]{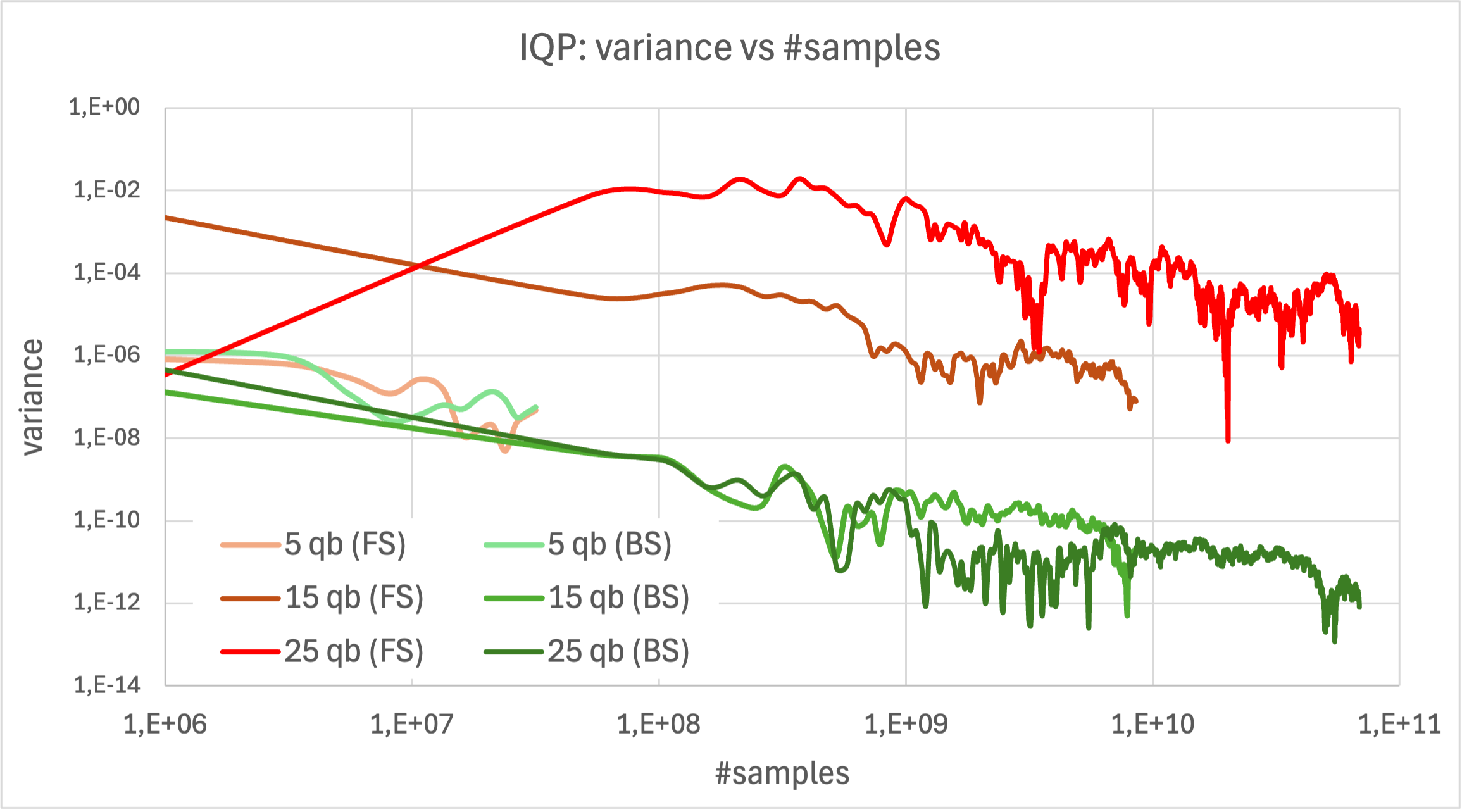}
    \caption{IQP 5, 15 and 25 qubits}
    \end{subfigure}
\\
\begin{subfigure}{\columnwidth}
    \includegraphics[width=\textwidth]{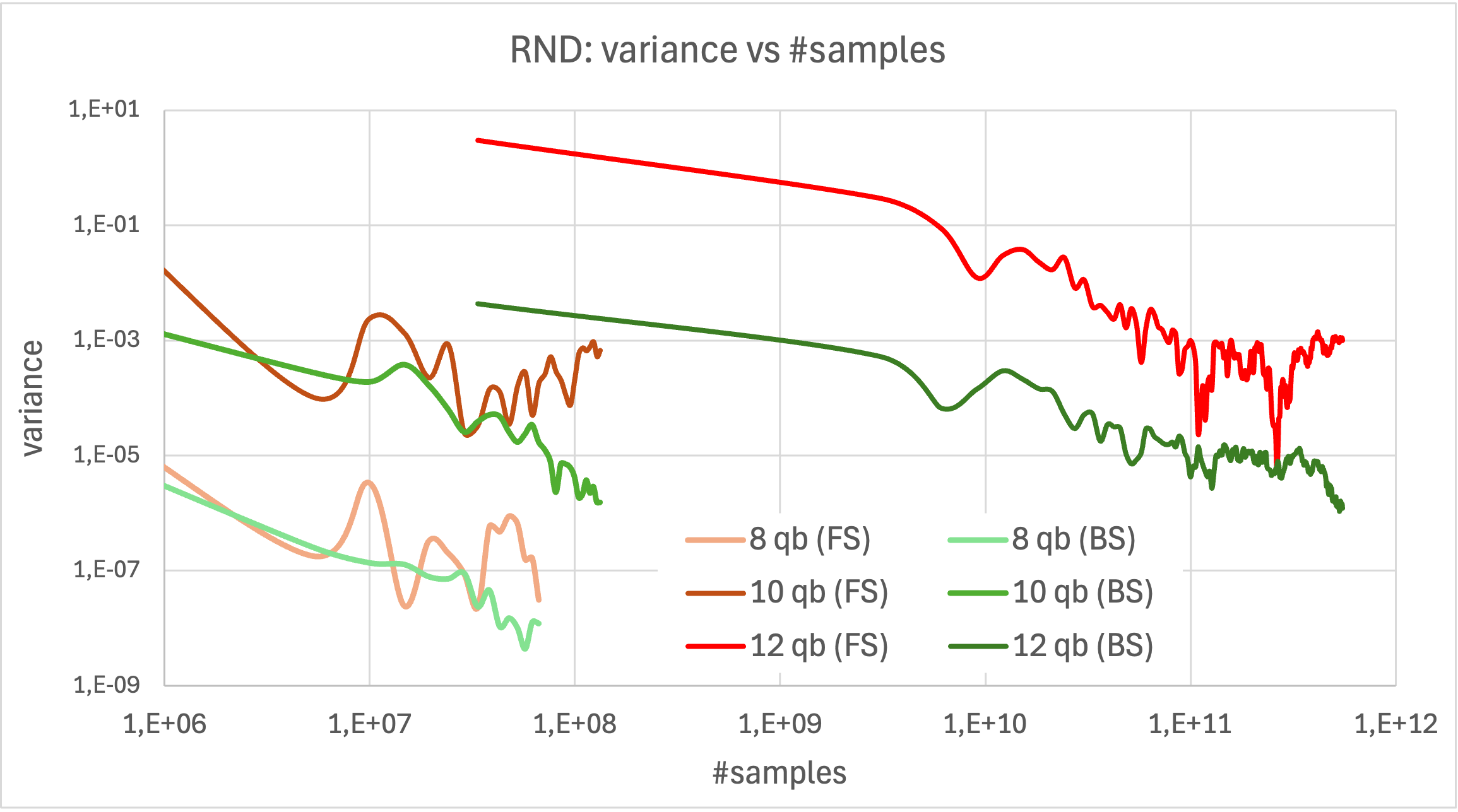}
    \caption{RND 8, 10 and 12 qubits}
    \end{subfigure}
\caption{Variance as a function of the number of Monte Carlo samples}
    \label{fig:VarVsSamples}
\end{figure}

Figure \ref{fig:VarVsSamples} illustrates how variance varies with the number of samples, $N$. {\bf BS} outperforms {\bf FS} for similar sampling counts, and this trend gets more significant as circuits get larger. For the \textbf{RND12} and \textbf{IQP25} circuits, {\bf BS }'s variance is smaller than {\bf FS} 's by 2 and 5 orders of magnitude, respectively. It must be kept in mind, however, that each {\bf BS} sample gives rise to $L$ paths.

\begin{figure}[htb]
    \centering
\begin{subfigure}{\columnwidth}
    \includegraphics[width=\textwidth]{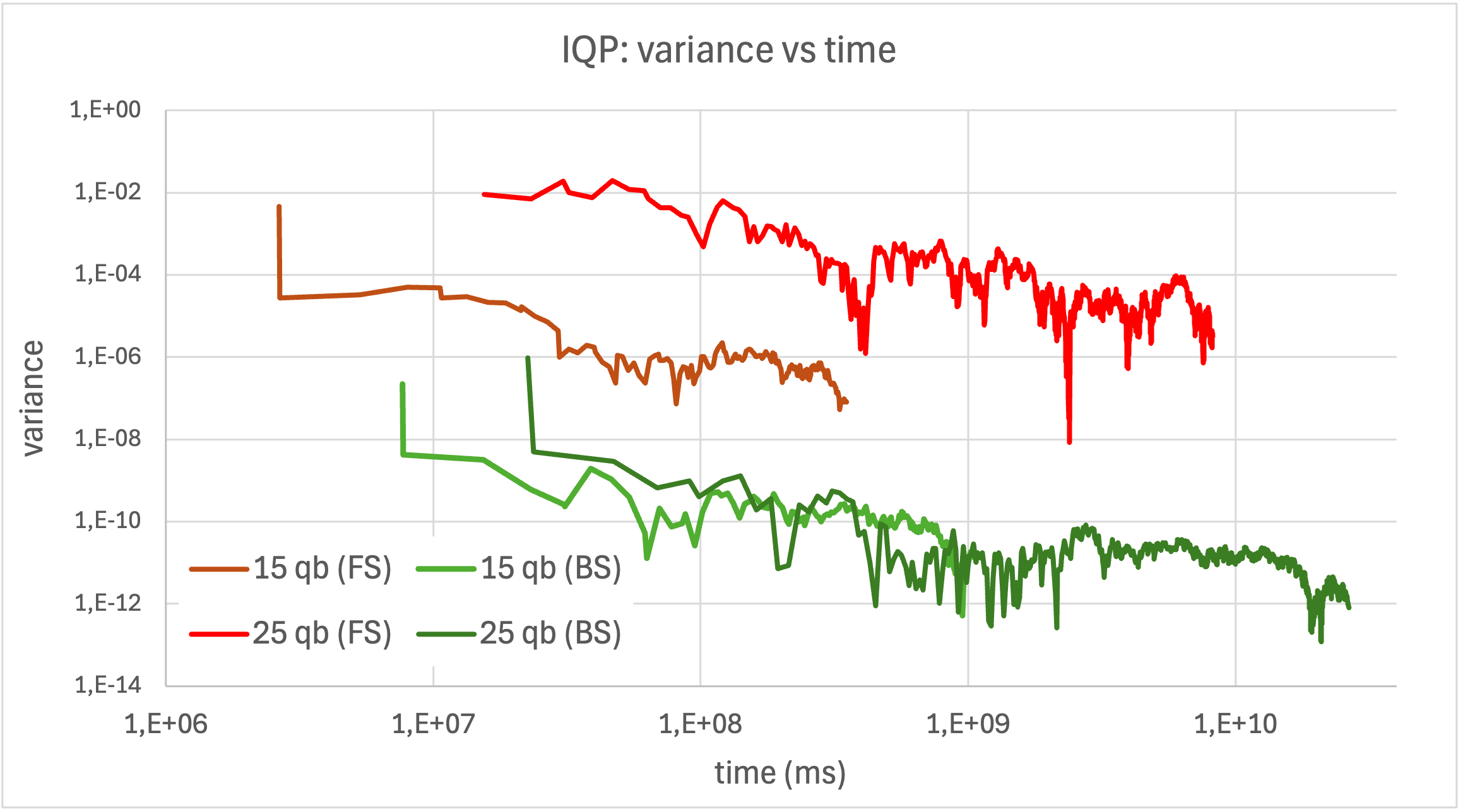}
    \caption{IQP 15 and 25 qubits}
    \end{subfigure}
\\
\begin{subfigure}{\columnwidth}
    \includegraphics[width=\textwidth]{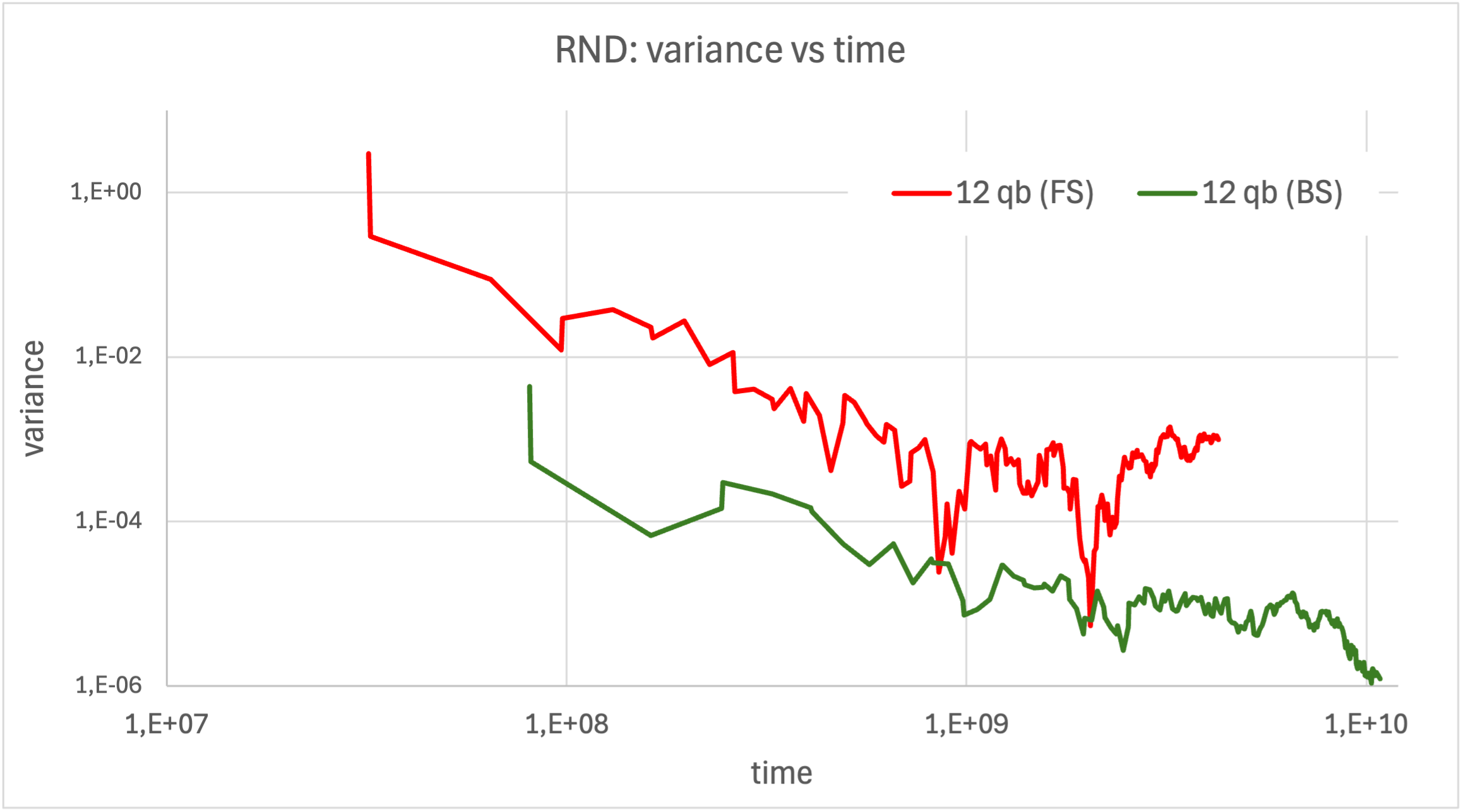}
    \caption{RND 12 qubits}
    \end{subfigure}
     \caption{Variance as a function of time}
     \label{fig:VarVsTime}
\end{figure}

Figure \ref{fig:VarVsTime} shows that, for the same execution time, {\bf BS} consistently outperforms {\bf FS}, as initially hypothesized. For one sample, {\bf FS} evaluates one forward path, whereas {\bf BS} assesses a pair of (forward, backward) connections corresponding to $L$ paths. Assessing these $L$ paths is quite efficient, resulting in variance gains ranging from one to six orders of magnitude as the circuits' sizes increase.

\begin{figure}[htb]
    \centering
    \begin{subfigure}{0.9\columnwidth}
    \includegraphics[width=\textwidth]{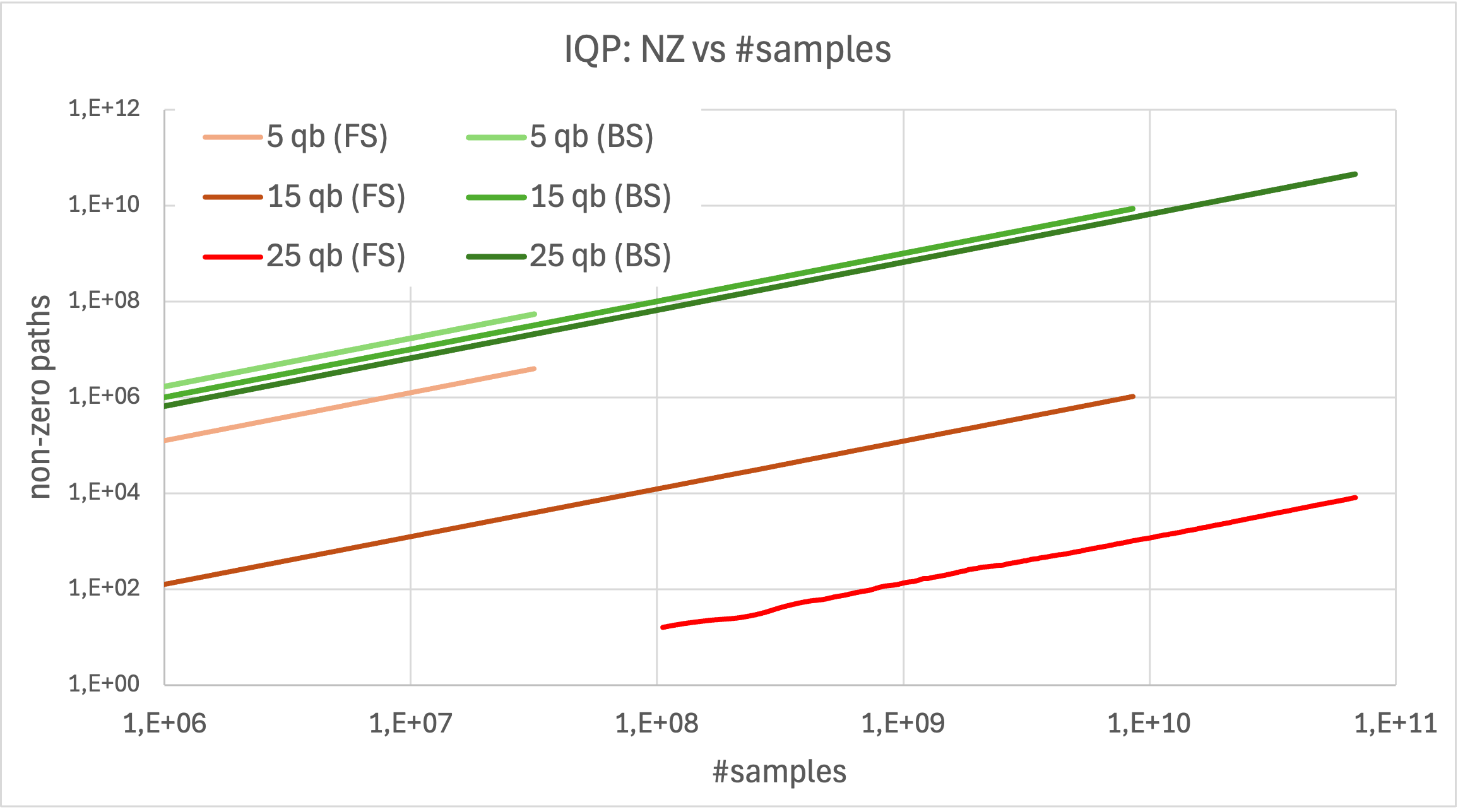}
    \caption{IQP 5, 15 and 25 qubits}
    \end{subfigure}
    \begin{subfigure}{0.9\columnwidth}
    \includegraphics[width=\textwidth]{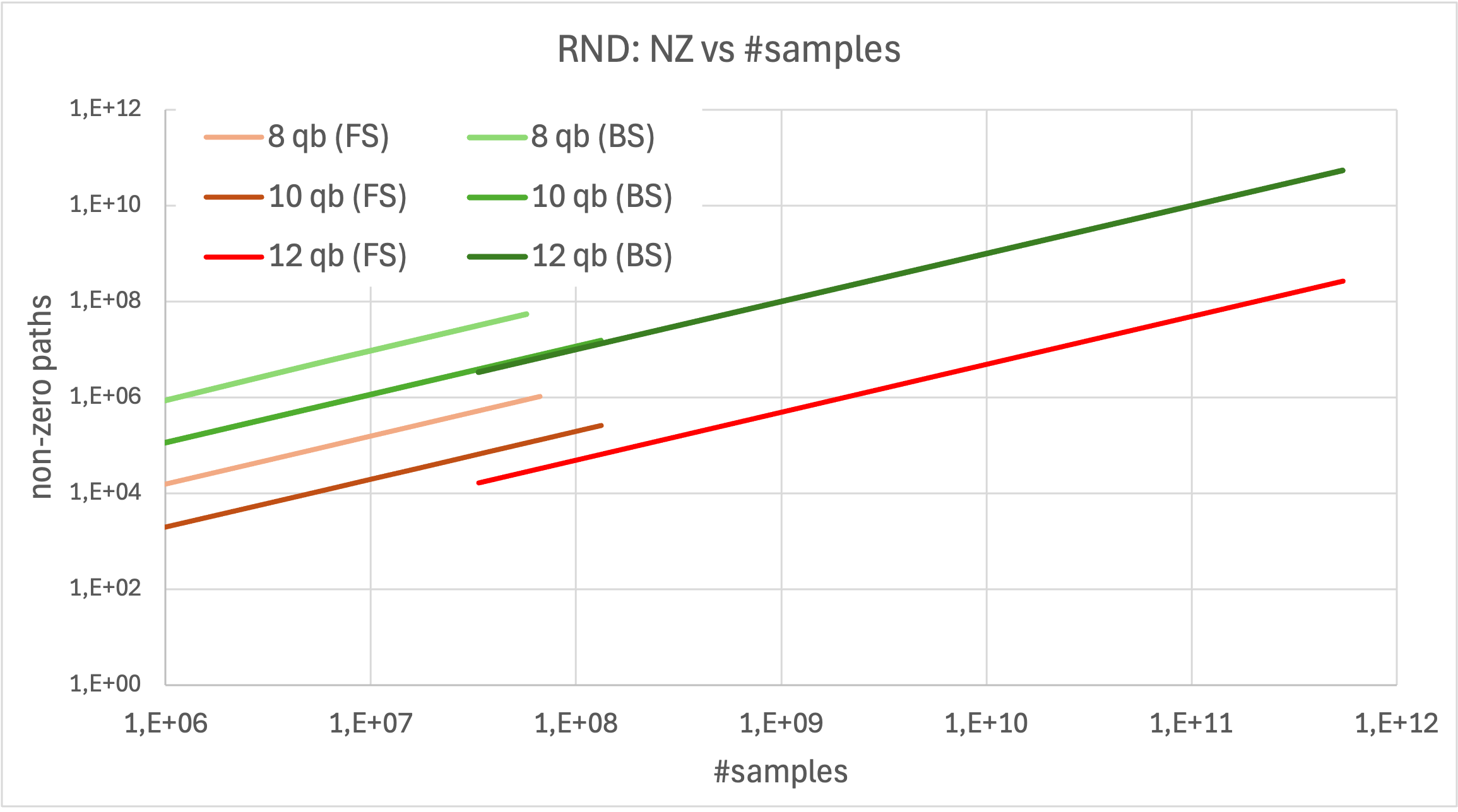}
    \caption{RND 8, 10 and 12 qubits}
    \end{subfigure}
    \caption{Number of non-zero amplitude paths as a function of the number of samples}
     \label{fig:n_NZ_Paths_vs_samples}
\end{figure}

{\bf BS} achieves superior convergence because it is more successful than {\bf FS} in the
discovery of non-zero amplitude paths, as illustrated in figure \ref{fig:n_NZ_Paths_vs_samples}. For \textbf{IQP25} and the same number of samples, {\bf BS} finds around six orders of magnitude more non-zero paths than {\bf FS}, whereas for \textbf{RND12} the gain is about two orders of magnitude. The bidirectional approach, by connecting forward and backward generated paths, is much more successful at building viable paths. Using \textbf{MIS} to evaluate the weights ensures good convergence. 

\subsection{H2 -- Simulation of large circuits}
\label{subsec:VLCircuits}

The motivation for Feynman sum of paths algorithms is their space complexity, which is $\mathcal{O}(nL)$, opening up the possibility of simulating very large circuits. Table \ref{tab:MemScaling} compares the memory footprint of the state vector (Qiskit 2.3.0) and the path summing simulators for \textbf{IQP} circuits (section \ref{subsec:MethodH1}) with varying numbers of qubits. The exponential space complexity of the former versus the linear complexity of the latter is well demonstrated. 

\begin{table}[htb]
    \centering
    \begin{tabular}{c|rr}
    
    \multicolumn{3}{c}{{\bf RSS} MiB} \\ \hline
    $n$ & \multicolumn{1}{c}{State vector} & \multicolumn{1}{c}{Feynman} \\ \hline
    5 & 109.0 & 1.2 \\
    15 & 131.0 & 1.8 \\
    20 & 217.0 & 2.0 \\
    25 & 2595.0 & 2.0 \\
    27 & 10267.0 & 2.4 \\ 
    35 & \multicolumn{1}{c}{--} & 2.5 \\
    45 & \multicolumn{1}{c}{--} & 2.8 \\ \hline
    \end{tabular}
\caption{{\bf RSS} (resident set size) in MiB for \textbf{IQP} circuits with varying number of qubits, using the state vector and the Feynman sum of paths simulators.}
\label{tab:MemScaling}
\end{table}

    

Table \ref{tab:HSC_Sim} presents the results of estimating transition amplitudes for the Hidden Shift Circuits (\textbf{HSC}) described in table \ref{tab:circuitsHSC}. These have 256, 1024 and 4096 qubits, which would be impossible to process with a naive Schrödinger-type simulator with exponential memory requirements. 4, 8, and 16 computing nodes, with 128 physical processing cores each, have been used to simulate the three circuits, respectively. The number of samples, i.e., pairs of forward and backward path segments, is $2^{24}$.  

The data show that {\bf BS} is capable of estimating the desired transition amplitude by finding some non-zero paths. These are very sparse path spaces; the {\bf FS} algorithm did not find any non-zero path for up to $2^{26}$ samples.  

The capability to simulate a 4096-qubit circuit is quite remarkable. A Schrödinger-based approach would require memory proportional to $\mathcal{O}(2^{4096})$, which is unfeasible.

\textbf{HSC} have been shown to be efficiently simulable using classical resources \cite{Amy2024}. Our empirical results' observed execution time and resource usage, presented in Table \ref{tab:HSC_Sim}, are consistent with polynomial, rather than exponential, scaling over the tested range.

%
    

\begin{table}[htb]
    \centering
    \begin{tabular}{ccccc}
     & \textbf{computing} & \textbf{time} &  &  \\ 
    \textbf{\#qb} & \textbf{nodes} & \textbf{(m:s)} & \textbf{\#NZ} & \textbf{Var} \\ \hline
    256 & 4 & 6m51s & 511 & $4e(-6)$ \\ \hline
    1024 & 8 & 13m56s & 502 & $5e(-4)$ \\ \hline
    4096 & 16 & 29m31s & 501 & $4e(-3)$ \\ \hline
    \end{tabular}
\caption{{\bf BS} simulation of very large \textbf{HSC} -- $N=2^{24}$}
\label{tab:HSC_Sim}
\end{table}

\subsection{H3 -- Sparsity and numerical sign problem} 

\begin{figure}[htb]
    \centering
    \begin{subfigure}{0.7\columnwidth}
    \includegraphics[width=\textwidth]{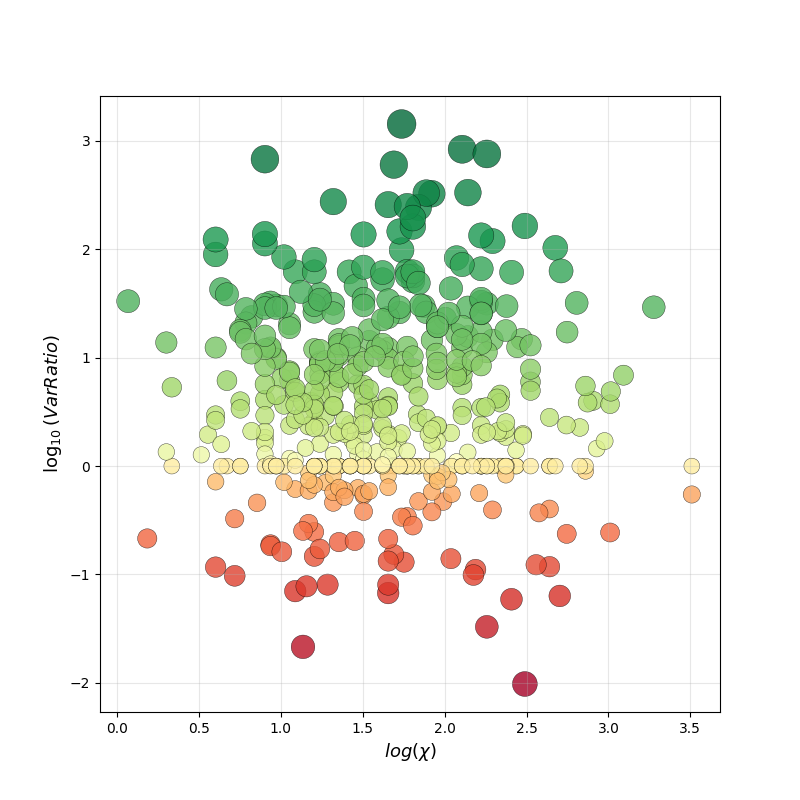}
    \caption{Variance ratio versus $\chi$}
    \end{subfigure} \\
    \begin{subfigure}{0.7\columnwidth}
    \includegraphics[width=\textwidth]{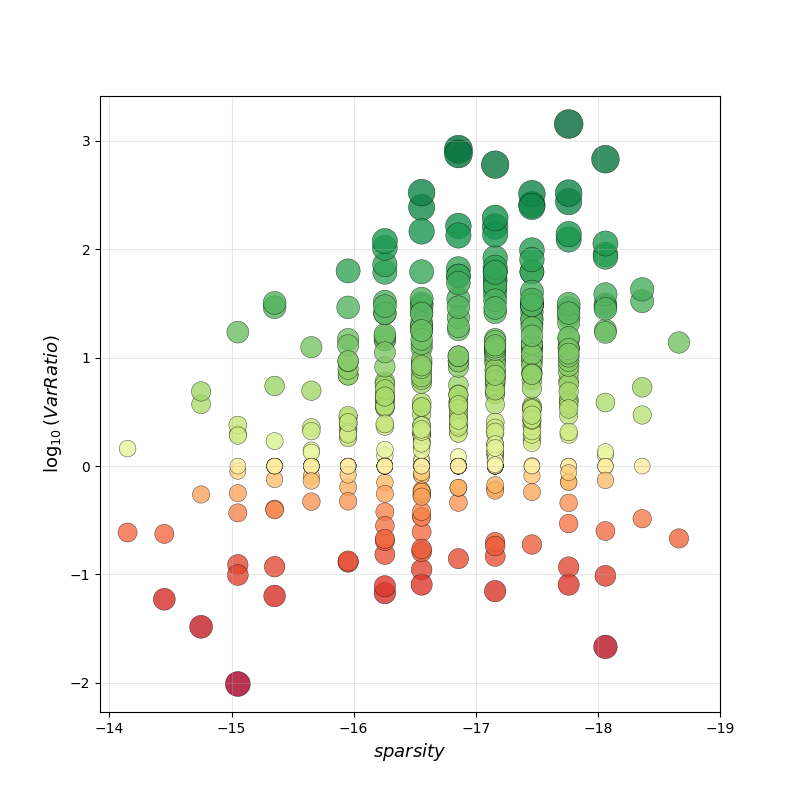}
    \caption{Variance ratio versus sparsity}
    \end{subfigure}
    \caption{Variance ratio as a function of sparsity and average phase - logarithmic scales}
     \label{fig:SparsityVsPhase}
\end{figure}

Figure \ref{fig:SparsityVsPhase} presents the logarithm of the ratio $\frac{Var_{FS}}{Var_{BS}}$ as a function of the logarithms of $\chi$ and sparsity, for all 500 circuits described in section \ref{subsec:MethodH3} and for constant execution time. Sparsity is defined as the logarithm of the ratio of the number of non-zero paths over the total number of paths -- more negative values indicate sparser path spaces.

The vertical axis indicates how many orders of magnitude is $Var_{BS}$ smaller than $Var_{FS}$ -- negative values in the vertical axis indicate that the {\bf FS} algorithm achieved smaller variance than  {\bf BS}, which happens for a reduced number of circuits. 

The variance ratio's logarithm for each circuit is displayed as a colored circle. Circles are yellow for unit ratios and get greener (respectively, redder) as the ratio increases (respectively, decreases). The circles' radii increase with the absolute value of the ratio logarithm.

Visual inspection confirms that the variance ratio is significantly less sensitive to $\chi$ than to sparsity. Pearson's coefficients of the variance ratio relative to $\chi$ and to sparsity are $\approx -0.02516$ and $\approx -0.13890$, respectively. These results are consistent with the hypothesis that the advantage of bidirectional multiple importance sampling, over forward sampling, comes from its capability to more effectively find and weight non-zero paths, rather than by handling the numerical sign problem caused by destructive interference among paths.

\section{Concluding Remarks}
\label{sec:conclusions}

This paper proposes a bidirectional multiple importance sampling Monte Carlo estimator for transition amplitudes $\langle \Psi_f | \hat{U} | \Psi_0 \rangle$, where $\hat{U}$ represents a quantum circuit. This estimator, denoted by $\tilde{I}_{BS}$, is inspired by the Bidirectional Path Tracing algorithm \cite{Veach1997} and its convergence rate is experimentally compared to that of a forward-only, unidirectional estimator, denoted by $\tilde{I}_{FS}$.

Empirical data on the convergence of the proposed estimator were collected using a wide range of quantum circuits and assuming noise-free, ideal conditions. These data confirm the superiority of the bidirectional algorithm, corroborating the three hypotheses proposed in the introduction.
\begin{description}
    \item[H1 -- Convergence rate: ] the variance of the bidirectional estimator is shown to decrease faster than that of the unidirectional estimator, as both a function of the number of samples and execution time;
    \item[H2 -- Simulation of large circuits: ] the $\mathcal{O}(nL)$ memory requirements of the PIMC approach compound with the bidirectional algorithm's efficient path sampling and weighting to allow the simulation of very large quantum circuits (estimation of transition amplitudes for 4096 qubits' circuits demonstrated in this work); 
    \item[H3 -- Sparsity and numerical sign problem: ] the demonstrated convergence rate improvements, over the unidirectional approach, are due to a more efficient sampling of the sparse path space, rather than to a better handling of destructive interference, i.e., MC's numerical sign problem.
\end{description}

\subsection{Future work}

This work focused on ideal, non-noisy systems. Future work entails addressing real, noisy systems, understanding under which conditions does reversibility holds and whether bidirectional sampling maintains its superior convergence properties.

\begin{figure}[htb]
    \centering
    \includegraphics[width=0.9\columnwidth]{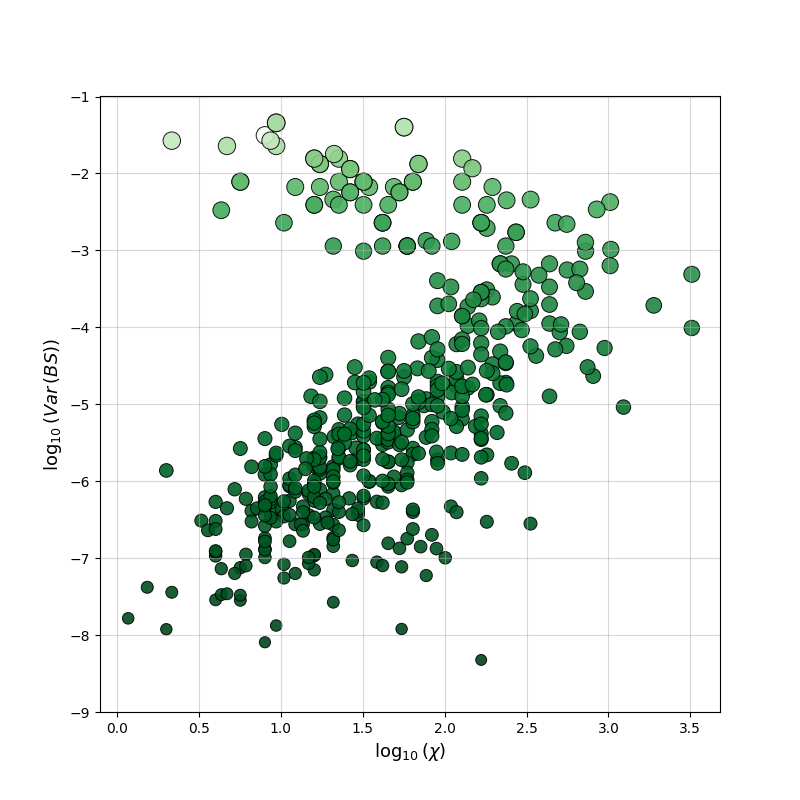}
    \caption{$Var_{\tilde{I}_{BS}}$ as a function of inverse squared average phase ($\chi$); each data point corresponds to the execution of each of the circuits described in section \ref{subsec:MethodH3} and for a constant number of samples - logarithmic scales}
     \label{fig:logVarBSvsChi}
\end{figure}

It has been shown that bidirectional multiple importance sampling of Feynman paths achieves better convergence rates than unidirectional sampling, being able to effectivelly sample within a very sparse path space.  However, the main challenge of PIMC algorithms is the numerical sign problem. As destructive interference among paths' amplitudes becomes more significant, the sampling requirements grow exponentially. Figure \ref{fig:logVarBSvsChi} shows how the variance of the bidirectional estimator increases with the inverse squared average phase ($\chi$) for a constant number of samples and the same 500 random circuits described in section \ref{subsec:MethodH3}.

\begin{figure}[htb]
    \centering
    \begin{subfigure}{0.7\columnwidth}
    \includegraphics[width=\textwidth]{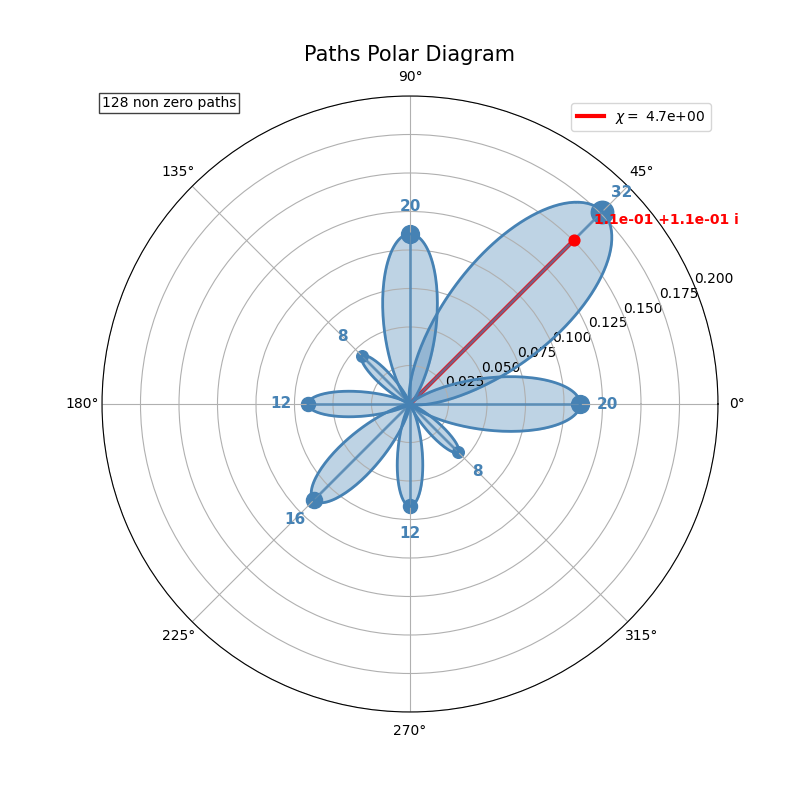}
    \caption{$\chi$ = 4.7}
    \label{fig:PolarPhases-smallX}
    \end{subfigure} \\
    \begin{subfigure}{0.7\columnwidth}
    \includegraphics[width=\textwidth]{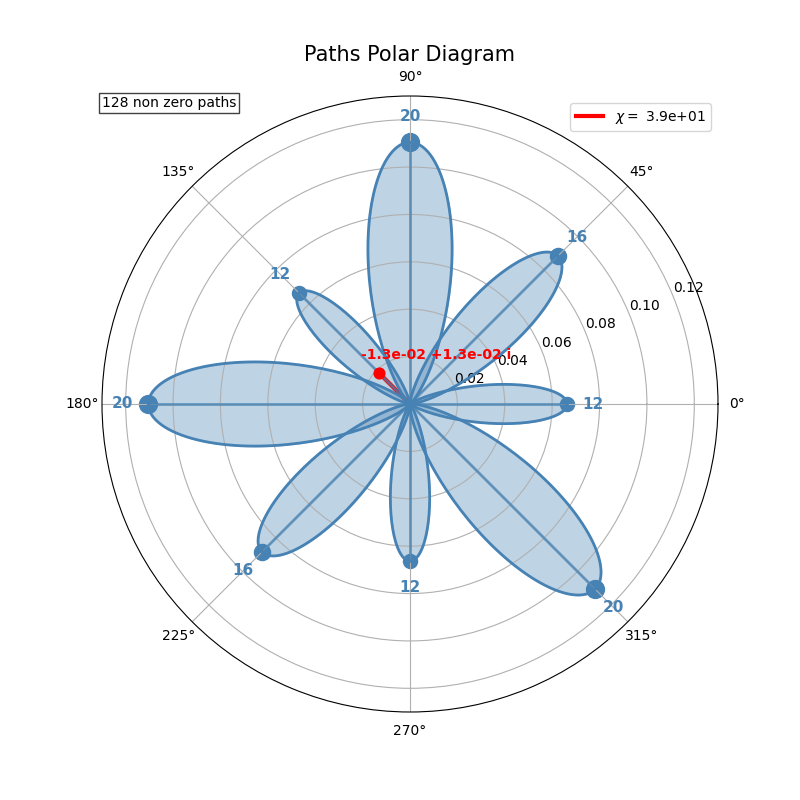}
    \caption{$\chi$ = 38.8}
    \label{fig:PolarPhases-largeX}
    \end{subfigure}
    \caption{Polar diagrams for circuits with 128 non zero paths and different inverse squared average phases ($\chi$)}
     \label{fig:PolarPhases}
\end{figure}

Figure \ref{fig:PolarPhases} illustrates how $\chi$ captures the strength of the numerical sign problem. Two random circuits, with exactly 128 non-zero paths in the path space associated with $\langle 0 | \hat{U} | 0 \rangle$ but exhibiting different $\chi$, were selected among those described in section \ref{subsec:MethodH3}.

The width of each lobe is proportional to the number of non-zero paths with that particular phase (this number is displayed next to the corresponding marker). The length of the lobe is equal to the sum of the magnitudes of all paths with the same phase. The $\langle 0 | \hat{U} | 0 \rangle$ transition amplitude, depicted in red, is given by the sum of the contributions of all paths.
Figure \ref{fig:PolarPhases-smallX} illustrates the case for $\chi=4.7$. The lobes within the first quadrant concentrate most of the total amplitude. Destructive interference is not very strong and the Monte Carlo estimator exhibits low variance.  Figure \ref{fig:PolarPhases-largeX}, on the other hand, illustrates the case for larger $\chi$. Paths amplitudes are now distributed across all quadrants and destructive interference among paths dominates the whole process, leading to high Monte Carlo sampling requirements. 

The above analysis raises the following research question: can alternative path representations be used to alleviate the numerical sign problem? Paths in this work are sequences of intermediate states represented in the computational basis. Many alternative representations are possible, such as stabilizer states or quasiprobability decompositions, for example. Can different representations lead to different distributions of the paths amplitudes lobes in the polar phase diagram, hopefully reducing destructive interference? Studying how the path representation impacts on Monte Carlo sampling requirements can lead to a deeper understanding of the dynamics of path sum simulators. A related question is whether a metric can be found that suggests what representation to use for every particular problem. Computing, or estimating, this metric must be computationally efficient.

\section{Data Availability}
\label{sec:dataavailability}

The quantum circuits analysed during the current study are available in the \textbf{ExperimentalData} git repository: \url{https://github.com/luisps/ExperimentalData.git}. 

\section*{Authors Contributions}

LPS and TBR both contributed to the original idea and results analysis. LPS was a major contributor to writing the manuscript and code development. All authors read and approved the final manuscript.

\section*{Acknowledgements}

This study was funded by FCT - Fundação para a Ciência e Tecnologia, I.P., grant number UID/50014/2025 (\url{https://doi.org/10.54499/UID/50014/2025}). The funder played no role in study design, data collection, analysis and interpretation of data, or the writing of this manuscript. 

\bibliography{BD-PIMC}

@phdthesis{Veach1997,
  title={Robust Monte Carlo methods for light transport simulation},
  author={Veach, Eric},
  year={1997},
  school={Stanford University},
  url = {http://graphics.stanford.edu/papers/veach_thesis/}
}

@inproceedings{Veach1995,
author = {Veach, Eric and Guibas, Leonidas J.},
title = {Optimally combining sampling techniques for Monte Carlo rendering},
year = {1995},
isbn = {0897917014},
publisher = {Association for Computing Machinery},
address = {New York, NY, USA},
url = {https://doi.org/10.1145/218380.218498},
doi = {10.1145/218380.218498},
booktitle = {Proceedings of the 22nd Annual Conference on Computer Graphics and Interactive Techniques},
pages = {419–428},
numpages = {10},
series = {SIGGRAPH '95}
}

@article{Vidal2003,
  title={Efficient classical simulation of slightly entangled quantum computations},
  author={Vidal, Guifr{\'e}},
  journal={Physical Review Letters},
  volume={91},
  number={14},
  pages={147902},
  year={2003}
}

@article{Orus2014,
  title={A practical introduction to tensor networks: Matrix product states and projected entangled pair states},
  author={Or{\'u}s, Rom{\'a}n},
  journal={Annals of Physics},
  volume={349},
  pages={117--158},
  year={2014}
}

@article{Viamontes2009,
  title={Improving gate-level simulation of quantum circuits},
  author={Viamontes, George F. and Markov, Igor L. and Hayes, John P.},
  journal={Quantum Information Processing},
  volume={8},
  number={2-3},
  pages={109--127},
  year={2009}
}

@article{burgholzer2020,
  title={Advanced simulation of quantum computations},
  author={Burgholzer, Lukas and Wille, Robert},
  journal={IEEE Transactions on Computer-Aided Design of Integrated Circuits and Systems},
  volume={39},
  number={12},
  pages={3952--3965},
  year={2020}
}

@inproceedings{gottesman1998,
  title        = {{The Heisenberg representation of quantum computers}},
  author       = {D. Gottesman},
  year         = 1999,
  booktitle    = {Proc. of the XXII Int. Colloquium on Group Theoretical Methods in Physics},
  pages        = {32--43},
  note          =  {Long version at arXiv preprint quant-ph/9807006}
}

@article{Leone2022,
  title={Quantifying magic: resource theory of nonstabilizerness},
  author={Leone, Lorenzo et al.},
  journal={PRX Quantum},
  volume={3},
  pages={020331},
  year={2022}
}

@article{aaronson2004,
  title={Improved simulation of stabilizer circuits},
  author={Aaronson, Scott and Gottesman, Daniel},
  journal={Physical Review A},
  volume={70},
  number={5},
  pages={052328},
  year={2004}
}

@article{Garcia2012,
  title={Simulation of quantum circuits via stabilizer frames},
  author={Garcia, Hector J. and Markov, Igor L.},
  journal={IEEE Transactions on Computer-Aided Design},
  volume={31},
  number={12},
  pages={1837--1849},
  year={2012}
}

@article{Dawson2006,
  title={The Solovay-Kitaev algorithm},
  author={Dawson, Christopher M. and Nielsen, Michael A.},
  journal={Quantum Information \& Computation},
  volume={6},
  number={1},
  pages={81--95},
  year={2006}
}

@article{Temme2017,
  title={Classical simulation of quantum circuits by half-space Monte Carlo},
  author={Temme, Kristan and Bravyi, Sergey and Gambetta, Jay M.},
  journal={Physical Review Letters},
  volume={119},
  number={18},
  pages={180509},
  year={2017}
}

@article{Pashayan2015,
  title={Estimating outcome probabilities of quantum circuits using quasiprobabilities},
  author={Pashayan, Hakop and Wallman, Joel J. and Bartlett, Stephen D.},
  journal={Physical Review Letters},
  volume={115},
  number={7},
  pages={070501},
  year={2015}
}

@article{Greenbaum2015,
    author = "Greenbaum, Daniel",
    title = "{Introduction to Quantum Gate Set Tomography}",
    journal = "arXiv preprint",
    eprint = "1509.02921",
    archivePrefix = "arXiv",
    primaryClass = "quant-ph",
    month = "9",
    year = "2015"
}

@article{Prosen2007,
  title = {Operator space entanglement entropy in a transverse Ising chain},
  author = {Prosen, Toma\v{z} and Pi\v{z}orn, Iztok},
  journal = {Phys. Rev. A},
  volume = {76},
  issue = {3},
  pages = {032316},
  numpages = {5},
  year = {2007},
  month = {Sep},
  publisher = {American Physical Society},
  doi = {10.1103/PhysRevA.76.032316},
  url = {https://link.aps.org/doi/10.1103/PhysRevA.76.032316}
}

@article{Rall2019,
  title = {Simulation of qubit quantum circuits via Pauli propagation},
  author = {Rall, Patrick and Liang, Daniel and Cook, Jeremy and Kretschmer, William},
  journal = {Phys. Rev. A},
  volume = {99},
  issue = {6},
  pages = {062337},
  numpages = {10},
  year = {2019},
  month = {Jun},
  publisher = {American Physical Society},
  doi = {10.1103/PhysRevA.99.062337},
  url = {https://link.aps.org/doi/10.1103/PhysRevA.99.062337}
}

@article{Bernstein1997,
author = {Bernstein, Ethan and Vazirani, Umesh},
title = {Quantum Complexity Theory},
journal = {SIAM Journal on Computing},
volume = {26},
number = {5},
pages = {1411-1473},
year = {1997},
doi = {10.1137/S0097539796300921}
}

@book{Nielsen2010,
  title     = {Quantum Computation and Quantum Information},
  author    = {Nielsen, M and Chuang, I.},
  year      = 2010,
  publisher = "Cambridge Univ. Press",
  address   = "London"
}

@misc{Biamonte2017,
      title={Tensor Networks in a Nutshell}, 
      author={Jacob Biamonte and Ville Bergholm},
      year={2017},
      eprint={1708.00006},
      archivePrefix={arXiv},
      primaryClass={quant-ph},
      url={https://arxiv.org/abs/1708.00006}, 
}

@article{Markov2008,
author = {Markov, Igor L. and Shi, Yaoyun},
title = {Simulating Quantum Computation by Contracting Tensor Networks},
journal = {SIAM Journal on Computing},
volume = {38},
number = {3},
pages = {963-981},
year = {2008},
URL = {https://doi.org/10.1137/050644756}
}

@Inbook{Wille2023,
author="Wille, Robert
and Hillmich, Stefan
and Burgholzer, Lukas",
title="Decision Diagrams for Quantum Computing",
bookTitle="Design Automation of Quantum Computers",
year="2023",
publisher="Springer International Publishing",
address="Cham",
pages="1--23",
isbn="978-3-031-15699-1",
url="https://doi.org/10.1007/978-3-031-15699-1_1"
}

@misc{Hong2025,
      title={A Tensor Network based Decision Diagram for Representation of Quantum Circuits}, 
      author={Xin Hong and Xiangzhen Zhou and Sanjiang Li and Yuan Feng and Mingsheng Ying},
      year={2025},
      eprint={2009.02618},
      archivePrefix={arXiv},
      primaryClass={quant-ph},
      url={https://arxiv.org/abs/2009.02618}, 
}

@inproceedings{Aaronson2017,
author = {Aaronson, Scott and Chen, Lijie},
title = {Complexity-theoretic foundations of quantum supremacy experiments},
year = {2017},
isbn = {9783959770408},
publisher = {Schloss Dagstuhl--Leibniz-Zentrum fuer Informatik},
address = {Dagstuhl, DEU},
booktitle = {Proceedings of the 32nd Computational Complexity Conference},
articleno = {22},
numpages = {67},
location = {Riga, Latvia},
series = {CCC '17}
}

@INPROCEEDINGS{Burgholzer2021,
  author={Burgholzer, Lukas and Bauer, Hartwig and Wille, Robert},
  booktitle={2021 IEEE International Conference on Quantum Computing and Engineering (QCE)}, 
  title={Hybrid Schrödinger-Feynman Simulation of Quantum Circuits With Decision Diagrams}, 
  year={2021},
  volume={},
  number={},
  pages={199-206},
  doi={10.1109/QCE52317.2021.00037}
}

@article{Bravyi2019,
  url = {https://doi.org/10.22331/q-2019-09-02-181},
  title = {Simulation of quantum circuits by low-rank stabilizer decompositions},
  author = {Bravyi, Sergey and Browne, Dan and Calpin, Padraic and Campbell, Earl and Gosset, David and Howard, Mark},
  journal = {{Quantum}},
  issn = {2521-327X},
  volume = {3},
  pages = {181},
  month = sep,
  year = {2019}
}

@article{Ayral2025,
  url = {https://comptes-rendus.academie-sciences.fr/physique/item/10.5802/crphys.229.pdf},
  title = {Classical and quantum algorithms for many-body problems},
  author = {Ayral, Thomas},
  journal = {{Comptes Rendus Physique}},
  volume = {26},
  pages = {25--89},
  year = {2025}
}

@article{Ferris2012,
  title = {Perfect sampling with unitary tensor networks},
  author = {Ferris, Andrew J. and Vidal, Guifre},
  journal = {Phys. Rev. B},
  volume = {85},
  issue = {16},
  pages = {165146},
  numpages = {10},
  year = {2012},
  month = {Apr},
  publisher = {American Physical Society},
  doi = {10.1103/PhysRevB.85.165146},
  url = {https://link.aps.org/doi/10.1103/PhysRevB.85.165146}
}

@article{Boixo2018,
  title = {Characterizing quantum supremacy in near-term devices},
  author = {Boixo, S. and Isakov, S.V. and Smelyanskiy, V.N. et al.},
  journal = {Nature Phys},
  volume = {14},
  pages = {595-600},
  year = {2018},
  url = {https://doi.org/10.1038/s41567-018-0124-x}
}

@inbook{Sauer2001,
author = {T. Sauer},
title = {THE FEYNMAN PATH GOES MONTE CARLO},
booktitle = {Fluctuating Paths and Fields},
chapter = {},
year = {2001},
publisher = {World Scientific},
pages = {29-42},
doi = {10.1142/9789812811240_0003},
URL = {https://www.worldscientific.com/doi/abs/10.1142/9789812811240_0003}
}

@article{Lafortune1998,
author = {Lafortune, Eric and Willems, Yves},
year = {1998},
month = {01},
title = {Bi-Directional Path Tracing},
volume = {93},
journal = {Proceedings of Third International Conference on Computational Graphics and Visualization Techniques (Compugraphics'}
}

@article{Troyer2005,
  title = {Computational Complexity and Fundamental Limitations to Fermionic Quantum Monte Carlo Simulations},
  author = {Troyer, Matthias and Wiese, Uwe-Jens},
  journal = {Physical Review Letters},
  volume = {94},
  issue = {17},
  year = {2005},
  month = {May},
  publisher = {American Physical Society},
  doi = {10.1103/PhysRevLett.94.170201}
}

@article{Bremner2011,
  title={Classical simulation of commuting quantum computations implies collapse of the polynomial hierarchy},
  author={Bremner, Michael J and Jozsa, Richard and Shepherd, Dan J},
  journal={Proceedings of the Royal Society A: Mathematical, Physical and Engineering Sciences},
  volume={467},
  number={2126},
  pages={459--472},
  year={2011},
  publisher={The Royal Society},
  doi={10.1098/rspa.2010.0301}
}

@misc{Marshall2024,
  author       = {Simon C. Marshall and
                  Scott Aaronson and
                  Vedran Dunjko},
  title        = {Improved separation between quantum and classical computers for sampling
                  and functional tasks},
  year         = {2024},
  url          = {https://doi.org/10.48550/arXiv.2410.20935},
  doi          = {10.48550/ARXIV.2410.20935},
  note         = {Accessed: 2025-07-28}
}

@article{Arute2019,
  title={Quantum supremacy using a programmable superconducting processor},
  author={Arute, F. and Arya, K. and Babbush, R.},
  journal={Nature},
  volume={574},
  pages={505--510},
  year={2019},
  doi={10.1038/s41586-019-1666-5}
}

@article{Peres2023,
  doi = {10.22331/q-2023-10-03-1126},
  url = {https://doi.org/10.22331/q-2023-10-03-1126},
  title = {Quantum circuit compilation and hybrid computation using {P}auli-based computation},
  author = {Peres, Filipa C. R. and Galv{\~{a}}o, Ernesto F.},
  journal = {{Quantum}},
  issn = {2521-327X},
  publisher = {{Verein zur F{\"{o}}rderung des Open Access Publizierens in den Quantenwissenschaften}},
  volume = {7},
  pages = {1126},
  month = oct,
  year = {2023}
}

@article{Xu2022,
  title = {Fermion sign bounds theory in quantum Monte Carlo simulation},
  author = {Zhang, Xu and Pan, Gaopei and Xu, Xiao Yan and Meng, Zi Yang},
  journal = {Phys. Rev. B},
  volume = {106},
  issue = {3},
  year = {2022},
  month = {Jul},
  publisher = {American Physical Society},
  doi = {10.1103/PhysRevB.106.035121},
  url = {https://link.aps.org/doi/10.1103/PhysRevB.106.035121}
}

@article{Alexandru2018,
  title = {Finite-density Monte Carlo calculations on sign-optimized manifolds},
  author = {Alexandru, Andrei and Bedaque, Paulo F. and Lamm, Henry and Lawrence, Scott},
  journal = {Phys. Rev. D},
  volume = {97},
  issue = {9},
  year = {2018},
  month = {May},
  publisher = {American Physical Society},
  doi = {10.1103/PhysRevD.97.094510},
  url = {https://link.aps.org/doi/10.1103/PhysRevD.97.094510}
}

@article{Veitch2012,
  author = {Victor Veitch and Christopher Ferrie and David Gross and Joseph Emerson},
  title = {Negative quasi-probability as a resource for quantum computation},
  journal = {New Journal of Physics},
  volume = {14},
  pages = {113011},
  year = {2012},
  doi = {10.1088/1367-2630/14/11/113011}
}

@article{Howard2017,
  author = {Mark Howard and Earl T. Campbell},
  title = {Application of a Resource Theory for Magic States to Fault-Tolerant Quantum Computing},
  journal = {Physical Review Letters},
  volume = {118},
  number = {9},
  pages = {090501},
  year = {2017},
  doi = {10.1103/PhysRevLett.118.090501}
}

@misc{Amy2024,
      title={Polynomial-Time Classical Simulation of Hidden Shift Circuits via Confluent Rewriting of Symbolic Sums}, 
      author={Matthew Amy and Lucas Shigeru Stinchcombe},
      year={2024},
      eprint={2408.02778},
      archivePrefix={arXiv},
      primaryClass={quant-ph},
      url={https://arxiv.org/abs/2408.02778}, 
}

@article{Cirac2021,
  title = {Matrix product states and projected entangled pair states: Concepts, symmetries, theorems},
  author = {Cirac, J. Ignacio and P\'erez-Garc\'{\i}a, David and Schuch, Norbert and Verstraete, Frank},
  journal = {Rev. Mod. Phys.},
  volume = {93},
  issue = {4},
  pages = {045003},
  numpages = {65},
  year = {2021},
  month = {Dec},
  publisher = {American Physical Society},
  doi = {10.1103/RevModPhys.93.045003},
  url = {https://link.aps.org/doi/10.1103/RevModPhys.93.045003}
}

@article{Rudolph2026,
  title = {Pauli Propagation: A Computational Framework for Simulating Quantum Systems},
  author = {Rudolph, Manuel S. and Jones, Tyson and Teng, Yanting and Angrisani, Armando and Holmes, Zo\"e},
  journal = {PRX Quantum},
  volume = {7},
  issue = {3},
  pages = {032001},
  numpages = {34},
  year = {2026},
  month = {Aug},
  publisher = {American Physical Society},
  doi = {10.1103/6vd7-l9bn},
  url = {https://link.aps.org/doi/10.1103/6vd7-l9bn}
}

@book{Kalos1964,
  title = {Monte Carlo Methods},
  author = {Mdvin H. Kalos and Paula A. Whitlock},
  year = {1964},
  publisher = {WILEY-VCH Verlag GmbH}
}

@article{Alexandru2022,
  title = {Complex paths around the sign problem},
  author = {Alexandru, Andrei and Ba\ifmmode \mbox{\c{s}}\else \c{s}\fi{}ar, G\"ok\ifmmode \mbox{\c{c}}\else \c{c}\fi{}e and Bedaque, Paulo F. and Warrington, Neill C.},
  journal = {Rev. Mod. Phys.},
  volume = {94},
  issue = {1},
  pages = {015006},
  numpages = {30},
  year = {2022},
  month = {Mar},
  publisher = {American Physical Society},
  doi = {10.1103/RevModPhys.94.015006},
  url = {https://link.aps.org/doi/10.1103/RevModPhys.94.015006}
}
\bibliographystyle{plain}

\end{document}